\documentclass[twocolumn]{aastex702}
\usepackage{natbib}

\graphicspath{{./}{figures/}}
\usepackage{color,subfigure,lineno,booktabs} 

\definecolor{firebrick}{rgb}{0.7, 0.13, 0.13}

\newcommand{\Hb}{H$\beta$}
\newcommand{\Ha}{H$\alpha$}

\def\OIII{[O\,{\sc iii}]}

\usepackage{pifont}

\shorttitle{NEXUS: A Transient at $z=5$} %
\shortauthors{Zhuang et~al.}

\begin{document}

\title{NEXUS: Discovery of a Luminous Young Transient at $z=5$}


\author[0000-0001-5105-2837]{Ming-Yang Zhuang}
\affiliation{Department of Astronomy, University of Illinois Urbana-Champaign, Urbana, IL 61801, USA}
\email[show]{mingyang@illinois.edu}

\author[0000-0003-1659-7035]{Yue Shen}
\affiliation{Department of Astronomy, University of Illinois Urbana-Champaign, Urbana, IL 61801, USA}
\affiliation{National Center for Supercomputing Applications, University of Illinois Urbana-Champaign, Urbana, IL 61801, USA}
\email{shenyue@illinois.edu}

\author[0000-0001-7201-1938]{Lei Hu}
\affiliation{Department of Physics and Astronomy, University of Pennsylvania, Philadelphia, PA 19104, USA}
\affiliation{McWilliams Center for Cosmology and Astrophysics, Department of Physics, Carnegie Mellon University, 5000 Forbes Avenue, Pittsburgh, PA 15213, USA}
\email{leihu@sas.upenn.edu}

\author[0000-0002-2361-7201]{Justin D. R. Pierel}
\affiliation{Space Telescope Science Institute, Baltimore, MD 21218, USA}
\altaffiliation{NASA Einstein Fellow}
\email{justin.pierel@gmail.com}

\author[0000-0003-0230-6436]{Zhiwei Pan}
\affiliation{Department of Astronomy, University of Illinois Urbana-Champaign, Urbana, IL 61801, USA}
\email{zhiweip@illinois.edu}

\author[0000-0002-8501-3518]{Zachary Stone}
\affiliation{Department of Astronomy, University of Illinois at Urbana-Champaign, Urbana, IL 61801, USA}
\email{stone28@illinois.edu}

\author[0000-0002-6523-9536]{Adam J.\ Burgasser}
\affiliation{Department of Astronomy \& Astrophysics, UC San Diego, La Jolla, CA 92093, USA}
\email{aburgasser@ucsd.edu}

\author[0000-0003-3310-0131]{Xiaohui Fan}
\affiliation{Steward Observatory, University of Arizona, 933 N. Cherry Ave., Tucson, AZ 85721, USA}
\email{xfan@arizona.edu}

\author[0000-0002-5612-3427]{Jenny E. Greene}
\affiliation{Department of Astrophysical Sciences, 4 Ivy Lane, Princeton University, Princeton, NJ 08540}
\email{}

\author[0000-0002-1605-915X]{Junyao Li}
\affiliation{Department of Astronomy, University of Illinois at Urbana-Champaign, Urbana, IL 61801, USA}
\email{junyaoli@illinois.edu}

\author[0000-0002-7633-431X]{Feige Wang}
\affiliation{Department of Astronomy, University of Michigan, 1085 S. University Ave., Ann Arbor, MI 48109, USA}
\email{fgwang@umich.edu}

\begin{abstract}
We report the discovery of AT~2026abck, a young transient in a star-forming galaxy at $z=5.017\pm0.001$, identified from Year-2 observations of the NEXUS JWST treasury program. AT~2026abck was detected on 2026 Aug 7 in NIRCam F200W imaging, with contemporaneous detection from parallel NIRCam F150W+F277W imaging. It brightened by $>1.4$~mag in F200W over $\lesssim 12$ rest-frame days, as inferred from the last non-detection. AT~2026abck has nearly identical magnitudes of $\sim27.4$~mag in F150W, F200W, and F277W, and is undetected in F444W ($>27.9$ mag, 3$\sigma$). It is the first robustly detected $z\gtrsim 5$ transient in a non-lensing field. A blackbody fit yields a hot photosphere ($\sim1.7\times10^4\, \mathrm{K}$) with a radius of $\sim7\times10^{14}$~cm ($\sim46$~au) and a bolometric luminosity of $\sim3\times10^{43}\, {\rm erg\, s^{-1}}$. Its off-nucleus location and hot, UV-bright SED favor a core-collapse supernova origin, while its temperature and luminosity exceed those of normal SNe~II at comparable early phases. We estimate that AT~2026abck was detected $\sim4\text{--}10$ rest-frame days after explosion. The host is an intermediate-mass ($M_*\approx10^{9.8}\,M_\odot$), star-forming (${\rm SFR}\approx20\,M_\odot\,{\rm yr^{-1}}$) galaxy on the $z\approx5$ main sequence, $\sim3.3$~dex more massive than the ultra-faint host of the lensed SN~Eos at $z=5.13$. Together, the two events show that CCSNe at $z>5$ occur across a wide range of host environments, and AT~2026abck demonstrates that high-cadence blank-field JWST surveys can capture high-redshift SNe during their early, UV-bright phase.
\end{abstract}

\section{Introduction}\label{sec:introduction}

Supernovae (SNe) mark the explosive deaths of stars and are among the most luminous transients in the Universe. Core-collapse supernovae (CCSNe) arise from massive stars ($\gtrsim8\,M_\odot$) at the end of their short lives, and therefore trace recent star formation. Through the injection of energy and newly synthesized elements into their surroundings, they also regulate the evolution of their host galaxies, an effect expected to be particularly strong in low-mass galaxies with shallow gravitational potentials \citep[e.g.,][]{Gentry2020, Gelli2024}. Type~Ia supernovae (SNe~Ia), the thermonuclear explosions of white dwarfs, serve as standardizable candles and led to the discovery of cosmic expansion acceleration \citep{Riess1998, Perlmutter1999}. CCSNe themselves are diverse, spanning hydrogen-rich SNe~II, stripped-envelope SNe~Ib/c, interaction-powered SNe~IIn, and rare superluminous SNe (SLSNe) \citep[e.g.,][]{Gal-Yam2019}. Their earliest phases are particularly informative. The first days after explosion are dominated by shock-cooling emission from the progenitor envelope and, in many cases, by interaction with dense circumstellar material (CSM), which together constrain the progenitor radius and its mass-loss history shortly before explosion \citep[e.g.,][]{NakarPiro2014, Forster2018, JacobsonGalan2024, Irani2024}.

SNe at high redshift offer a direct view of massive stars and their deaths in the early Universe. The CCSN rate as a function of redshift provides a probe of the cosmic star-formation history that is independent of the luminosity-to-SFR conversions and dust corrections required for galaxy-based measurements \citep{MadauDickinson2014, Strolger2015}. At $z\gtrsim3$, the lower metallicities and denser, burstier star-forming environments of young galaxies may alter the initial mass function (IMF), the ``explodability'' of massive stars, and their pre-explosion mass loss, all of which leave imprints on SN rates, subtype fractions, and explosion properties \citep[e.g.,][]{Ibeling2013, Pessi2023, SN_Eos_host}. The spectra of SNe~II in turn provide a direct probe of the metallicity of the gas from which their progenitors formed \citep{Dessart2014, Anderson2016, SN_Eos}. High-redshift SNe~Ia test for redshift-dependent luminosity evolution in a regime where the distance--redshift relation is relatively insensitive to dark energy \citep{RiessLivio2006, Pierel2025}, while the most luminous explosions, such as SLSNe and pair-instability SNe, may probe the first generations of very massive, metal-poor stars \citep[e.g.,][]{Kasen2011, Jeon2026}.

Until recently, observational studies of high-$z$ SNe were severely limited. Cosmological dimming and the redshifting of the rest-frame UV--optical emission into the infrared make SNe at $z\gtrsim2$ exceptionally difficult to detect. The $\sim900$-orbit CANDELS program with the Hubble Space Telescope discovered only a handful of SNe at $z\approx2$ \citep{Rodney2014}, and ground-based searches have identified a handful of SLSNe and luminous SNe~IIn out to $z\approx3.9$ \citep[e.g.,][]{Cooke2009, Cooke2012}. JWST has completely transformed this landscape. Multi-epoch imaging from blank-field surveys has yielded more than 200 SN candidates, with host redshifts extending to $z\approx6$ \citep{JADES_transients, COSMOS-Web_transients, Yan2026, NEXUS_transients}, together with lensed discoveries in cluster fields \citep[e.g.,][]{Frye2024, Yan2023}. Spectroscopic follow-up of a small subset of these candidates has confirmed SNe~Ia and CCSNe out to $z\approx3.6$ \citep{Pierel2024, Siebert2024, Coulter2026}. Beyond $z\approx4$, however, the census remains sparse. Most events are observed at only one or a few epochs, and few have been caught within days of explosion, when the emission most directly constrains the progenitor and its immediate environment. At $z>5$, the only spectroscopically confirmed SN is SN~Eos, a strongly lensed SN~IIP at $z=5.133$ whose spectroscopic and host characterization relying on lensing magnification \citep{SN_Eos, SN_Eos_host}.

In this paper, we report the discovery of AT~2026abck, a young transient at $z=5.017\pm0.001$ detected in a non-lensed field on 2026 August 7. The transient is hot and luminous, and its properties suggest a CCSN observed within $\sim4\text{--}10$ rest-frame days of explosion. Its host is an intermediate-mass, star-forming galaxy on the $z\approx5$ main-sequence. This paper is organized as follows. Section~\ref{sec:data} describes the observations and data reduction. Section~\ref{sec:det} presents the detection of AT~2026abck and its photometric properties. Section~\ref{sec:host} characterizes its host galaxy. Section~\ref{sec:discussion} discusses the physical nature of the transient, its explosion epoch, and its comparison with SN~Eos, and Section~\ref{sec:conclusion} summarizes our conclusions. Throughout, we adopt a flat $\Lambda$CDM cosmology with $H_0=70 \, {\rm km\, s^{-1}\, Mpc^{-1}}$, $\Omega_m=0.3$, and $\Omega_\Lambda=0.7$, and all magnitudes are in the AB system.

\begin{figure*}[ht]
    \centering
    \includegraphics[width=0.9\linewidth]{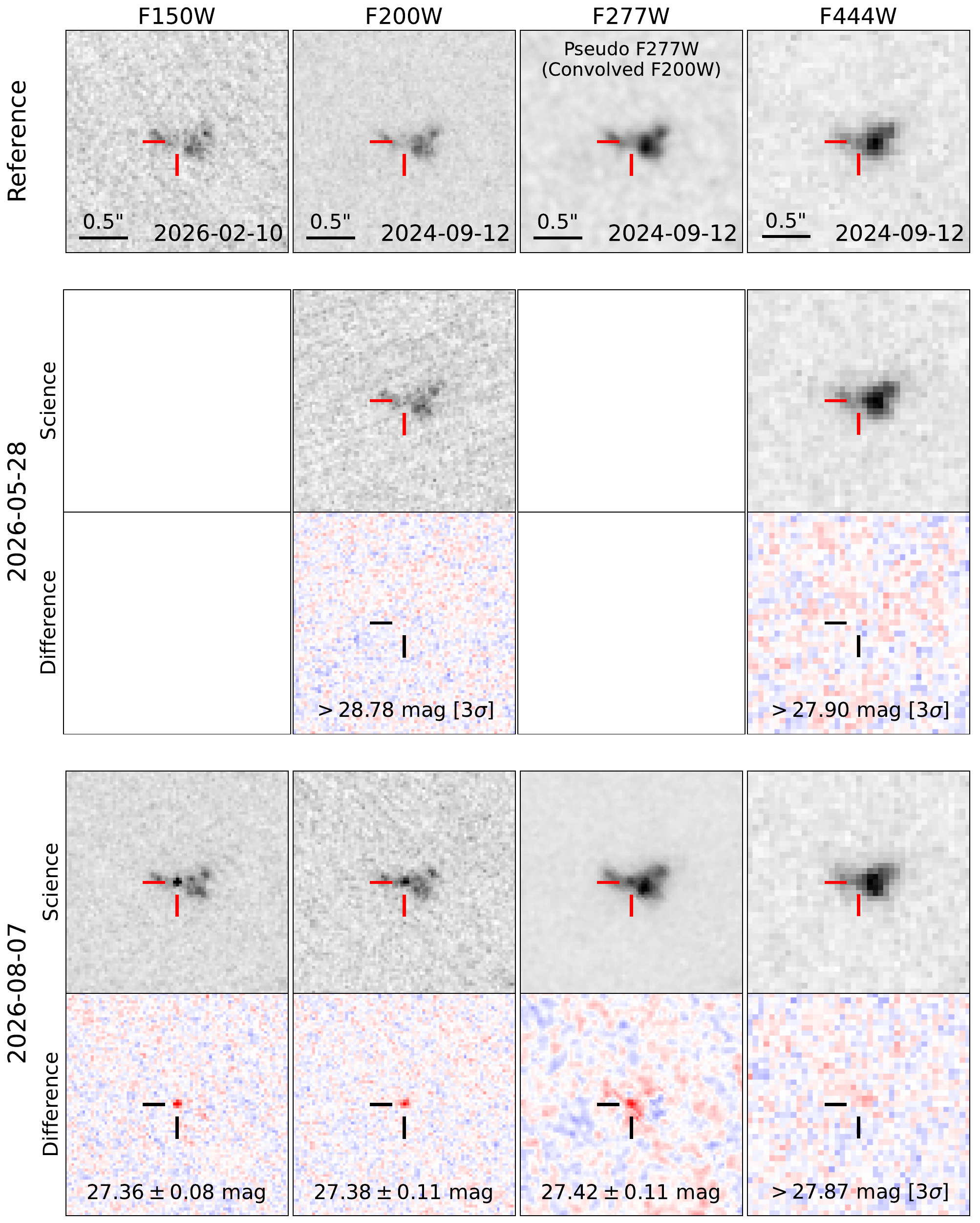}
    \caption{NIRCam discovery images of AT\,2026abck at $z=5.017$. Columns show F150W, F200W, F277W, and F444W bands, while rows from top to bottom show pre-explosion reference, Deep Ep07 science and difference images on 2026 May 28, Deep Ep08 science and difference images on 2026 Aug 7, respectively. The F277W reference is the F200W image convolved to the F277W PSF and rescaled to the host flux. PSF-photometry or $3\sigma$ upper limits are listed in the difference panels. AT~2026abck is indicated by red and black crosshairs in each panel. }
    \label{fig:discovery}
\end{figure*}

\begin{figure*}[t]
    \centering
    \includegraphics[width=\linewidth]{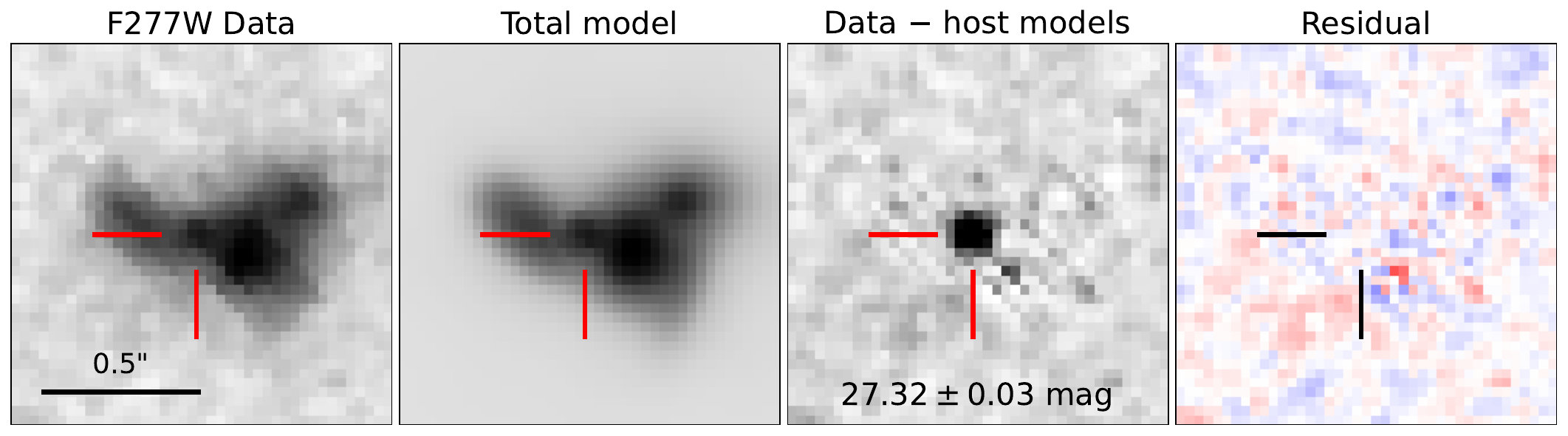}
    \caption{\texttt{galfitm} decomposition results of F277W image in Deep Ep08. Panels from left to right show the data, total model (transient + host), data minus all host models, and the residual after subtracting the total model. The position of AT~2026abck is marked with crosshairs. Magnitude of the transient is shown in the third panel. Note that \texttt{galfitm} underestimates the uncertainty as it does not consider noise correlation in drizzled images.}
    \label{fig:galfitm_results}
\end{figure*}

\begin{figure}[t]
    \centering
    \includegraphics[width=\linewidth]{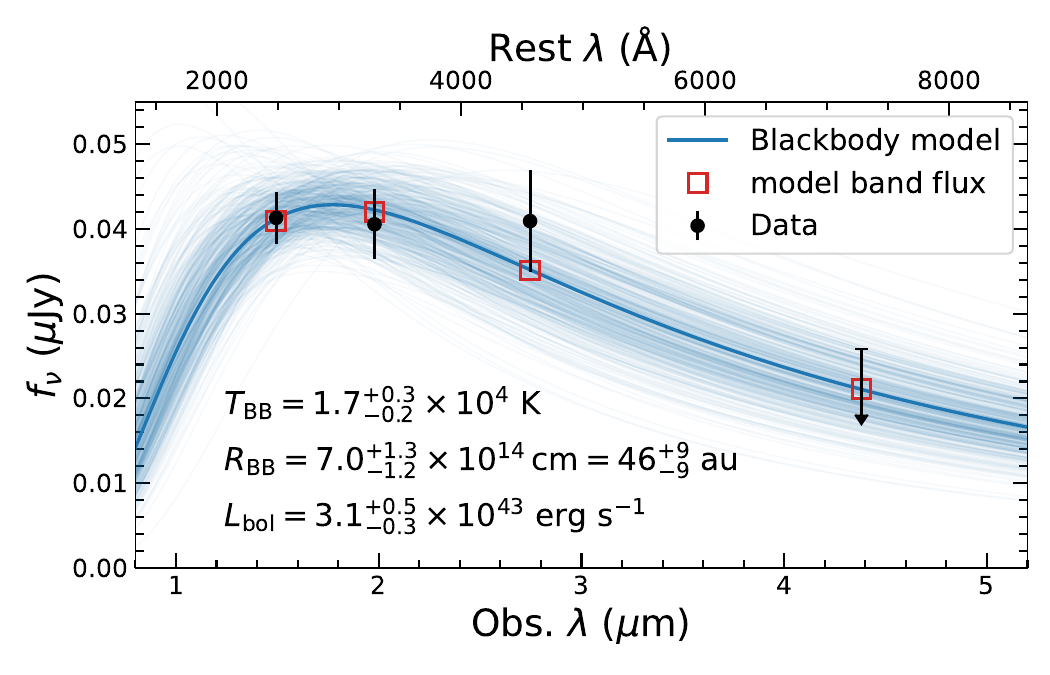}
    \caption{Blackbody fit to the discovery epoch SED (black dots) of AT~2026abck. Errorbars indicate $1\sigma$ uncertainties, with $3\sigma$ upper limit shown as downward arrow. The thick and faint blue curves represents the median model and models from 500 posterior draws, respectively. Red squares indicate the averaged model band fluxes. Best-fit parameters are listed in the lower left corner, with uncertainties from 16th--84th percentiles. $T_{\rm BB}$, $R_{\rm BB}$, and $L_{\rm bol}$ represent the effective temperature, radius of the sphere, and the bolometric luminosity of the blackbody model, respectively.}
    \label{fig:BB_fit_SED}
\end{figure}

\section{Observations and Data}\label{sec:data}

We use NIRCam imaging data from the NEXUS program \citep{NEXUS}. NEXUS has two overlapping tiers: the \textbf{Wide tier} covers a footprint of $\sim 400\,{\rm arcmin^2}$ with NIRCam imaging in six filters (F090W, F115W, F150W, F200W, F356W, F444W) and WFSS slitless spectroscopy in F322W2 and F444W; the \textbf{Deep tier} covers the central $\sim 50\ {\rm arcmin^2}$ within the Wide tier, and performs NIRSpec MSA/PRISM spectroscopy and F200W+F444W NIRCam imaging on a 2-month cadence through early 2028. The Wide tier is revisited annually over three cycles with a $\Delta t$ of about 17 months (PA orientation about $\pm 150$~deg). However, due to additional scheduling constraints, the first Wide epoch (Wide Ep01) was split in two, with the central $\sim 100\,{\rm arcmin^2}$ observed in September 2024 \citep{NEXUS_EDR} and the rest of Wide Ep01 observed in June 2025. For both Wide and Deep observations, there are coordinated parallel observations with MIRI imaging and NIRCam imaging in additional filters, but these parallel observations have non-uniform and smaller coverage than the primary observations.

We use the F200W+F444W from the central $\sim100\, {\rm arcmin^2}$ of the NEXUS-Wide~Ep01 NIRCam imaging as the reference and utilize the subsequent cadenced NIRCam F200W+F444W observations in NEXUS Deep \citep{NEXUS_QDR} to search for transients. Transient identification follows the procedure described in \citet{NEXUS_transients}, which we summarize briefly below. In each Deep epoch, the F200W and F444W mosaics are differenced against the Wide~Ep01 references on a common pixel grid with uniform photometric calibration, using the cross-convolution scheme of \citet{Hu2024}: The science and reference images are cross-convolved with each other's \texttt{stpsf} PSF models and subtracted. The resulting difference image is then decorrelated to whiten the background noise and recover a sharp effective PSF. Candidates are extracted with \textsc{SExtractor} \citep{SExtractor} and are matched within 0\farcs2. Dual-band detections enter the primary sample directly; single-band detections undergo further validation. Candidates detected in $\geq2$ individual Level~2 exposures are included in the primary sample. For single-dither coverage, inclusion requires a prior-epoch counterpart that was either already in the primary sample or previously flagged as a promising single-exposure candidate based on a plausible host association or $\mathrm{SNR} \gtrsim 15$. Source-injection tests give $50\%$ completeness at 28.15 and 27.29~mag in F200W and F444W.

\begin{deluxetable}{lccc}
\tablecaption{Photometry of AT~2026abck and its host galaxy\label{tab:fluxes}}
\tablewidth{0pt}
\tablehead{
\colhead{Filter} & \multicolumn{3}{c}{Flux Density (AB mag)}\\
\cline{2-4}
\colhead{Date} & \colhead{2026 May 28} & \colhead{2026 Aug 7} & \colhead{Host}
}
\startdata
F090W & \nodata & \nodata & $25.44 \pm 0.21$ \\
F115W & \nodata & \nodata & $25.29 \pm 0.16$ \\
F150W & \nodata & $27.36 \pm 0.08$ & $25.19 \pm 0.11$ \\
F200W & $>28.78$ & $27.38 \pm 0.11$ & $25.15 \pm 0.05$ \\
F277W & \nodata & $27.37 \pm 0.16$\tablenotemark{a} & $24.14 \pm 0.03$\tablenotemark{b} \\
F356W & \nodata & \nodata & $24.47 \pm 0.03$ \\
F444W & $>27.90$ & $>27.87$ & $24.33 \pm 0.02$ \\
\enddata
\tablecomments{Transient fluxes are measured on the difference image using \texttt{jwst\_psfmc}. Host fluxes are aperture photometry on the SN-free images, corrected to total flux using the empirical PSF encircled-energy curve of each band. Uncertainties are $1\sigma$ with non-detections shown as $3\sigma$ upper limits. All fluxes have been corrected for Galactic extinction.}
\tablenotetext{a}{We adopt the mean flux from difference image and \texttt{galfitm} with an uncertainty chosen conservatively to encompass the $\pm1\sigma$ ranges of both measurements.}
\tablenotetext{b}{F277W is the only band with no pre-explosion epoch, so the host flux is measured on the image obtained after subtracting the \texttt{galfitm} PSF component of AT~2026abck.}
\end{deluxetable}

\section{Transient Detection and Properties}\label{sec:det}

AT~2026abck was detected on 2026 August 7 in the Deep~Ep08 observations: in F200W from the primary NIRCam imaging, and in F150W and F277W from the parallel NIRCam imaging obtained alongside NIRSpec MSA/PRISM spectroscopy on the same day (Figure~\ref{fig:discovery}). It is undetected in both F200W and F444W in the Deep~Ep07 observations on 2026 May 28, which is 71 observed ($\sim11.8$ rest-frame) days earlier, with a $3\sigma$ limit of 28.78~mag in F200W. The transient therefore brightened by $>1.4$~mag in F200W between the two epochs.

We measure the photometry of AT~2026abck on the difference images using \texttt{jwst\_psfmc}\footnote{\url{https://github.com/mingyangzhuang/jwst_psfmc}} \citep{NEXUS_transients}, which accounts for correlated noise in the drizzled images to provide reliable uncertainties. Because F277W was not observed in Deep~Ep07, we construct a pseudo-reference image for F277W from the Deep~Ep07 F200W image. We convolve it with a kernel that homogenizes its point spread function (PSF) to that of F277W and rescale it to match the host flux in F277W. This approach assumes that the host morphology is similar in the two bands, which may not hold exactly because F277W contains the strong \OIII\ emission lines of the host at $z=5.017$. However, AT~2026abck lies between the two host components, where the host surface brightness in both the continuum and line emission is low, so any mismatch in the host model should have a limited effect on the transient flux. As an independent check, we also model the Deep~Ep08 F277W image directly with \texttt{galfitm} \citep{galfitm}, using a PSF for the transient and four S\'{e}rsic components for the host. The entire system is well reproduced, and the transient is clearly detected with $\sim27.3$~mag after decomposing the host (Figure~\ref{fig:galfitm_results}). This agrees with the difference-image measurement within the uncertainties, confirming the F277W detection. Since \texttt{galfitm} does not account for correlated noise and thus underestimates the uncertainty, we adopt the mean of the two flux measurements with an uncertainty chosen conservatively to encompass the $\pm1\sigma$ ranges of both measurements. The photometry is listed in Table~\ref{tab:fluxes}.

At the redshift of the host ($z=5.017\pm0.001$; Section~\ref{sec:host}), the detected bands sample rest-frame $\sim2500\text{--}4600$~\AA. We characterize the continuum by fitting a single-temperature blackbody to the Deep~Ep08 photometry using the Markov chain Monte Carlo sampler \texttt{emcee} \citep{emcee}, treating the F444W $3\sigma$ upper limit with the cumulative distribution function of the standard normal distribution. The observed flux density of a sphere of radius $R_{\rm BB}$ and effective temperature $T_{\rm BB}$ is
\begin{equation}
f_\lambda(\lambda) = \frac{\pi B_\lambda\!\left[\lambda/(1+z),\,T_{\rm BB}\right]}{1+z}\left(\frac{R_{\rm BB}}{D_L}\right)^{2},
\end{equation}
where $B_\lambda$ is the Planck function and $D_L$ is the luminosity distance. The fit gives $T_{\rm BB}=1.7^{+0.3}_{-0.2}\times10^{4}$~K and $R_{\rm BB}=7.0^{+1.3}_{-1.2}\times10^{14}$~cm ($46\pm9$~au), corresponding to bolometric luminosity $L_{\rm bol}=4\pi R_{\rm BB}^2\sigma T_{\rm BB}^4=3.1^{+0.5}_{-0.3}\times10^{43}$~erg~s$^{-1}$. The observed F150W magnitude corresponds to a rest-frame absolute magnitude of $M_{\rm UV}\approx-19.0$~mag at $\sim2500$~\AA. The nearly flat $f_\nu$ across F150W, F200W, and F277W places the SED peak near rest-frame $\sim3000$~\AA, while the F444W upper limit constrains the long-wavelength side of the SED. If the explosion occurred after the Deep~Ep07 epoch, the photosphere must have expanded at a mean velocity of $v\gtrsim R_{\rm BB}/(11.8~{\rm d})\approx7\times10^{3}\, {\rm km\, s^{-1}}$. We do not correct for host-galaxy extinction; adopting the host-integrated $A_V\approx0.2$~mag (Section~\ref{sec:host}) would increase $T_{\rm BB}$ and $L_{\rm bol}$. A single blackbody is also a simplification, as metal-line blanketing can suppress the rest-frame UV flux of SNe. We discuss the physical interpretation of these properties in Section~\ref{sec:transient_nature}.

\begin{figure}[t]
    \centering
    \includegraphics[width=\linewidth]{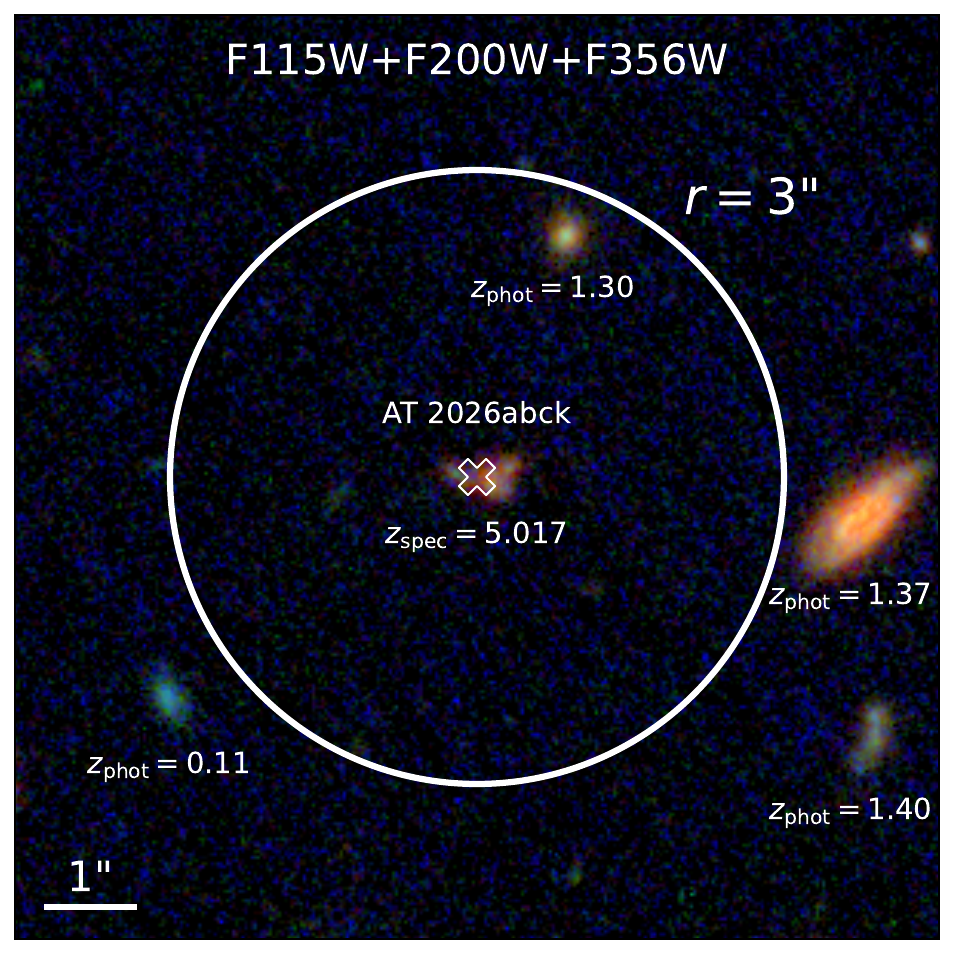}
    \caption{NIRCam F115W+F200W+F356W false-color stamp around AT~2026abck. The white cross marks the position of AT~2026abck and the white circle shows a 3\arcsec\ radius around it. Redshifts of nearby bright galaxies are labeled. Only two galaxies within $3\arcsec$-radius are brighter than 28 mag in F200W.}
    \label{fig:SN_environment}
\end{figure}

\begin{figure*}[t]
    \centering
    \includegraphics[width=0.47\linewidth]{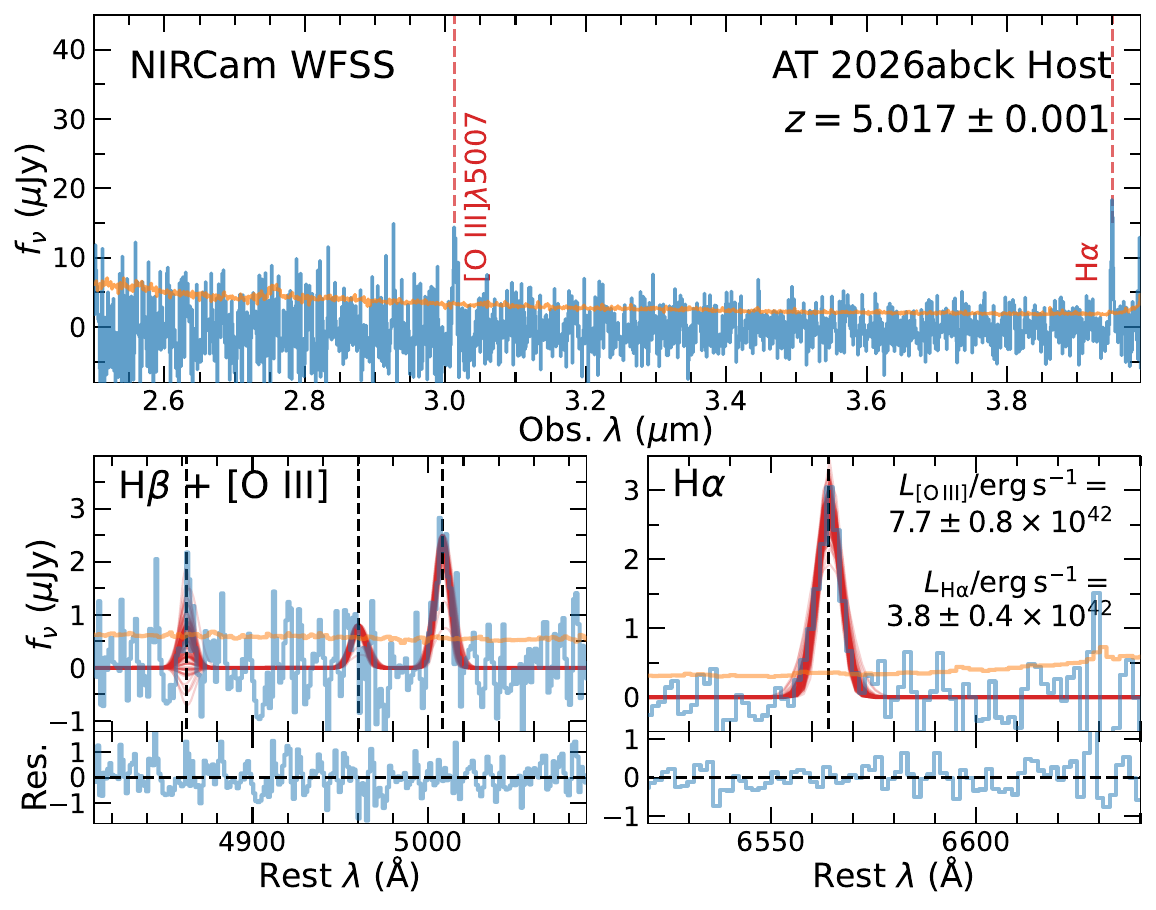}
    \includegraphics[width=0.5\linewidth]{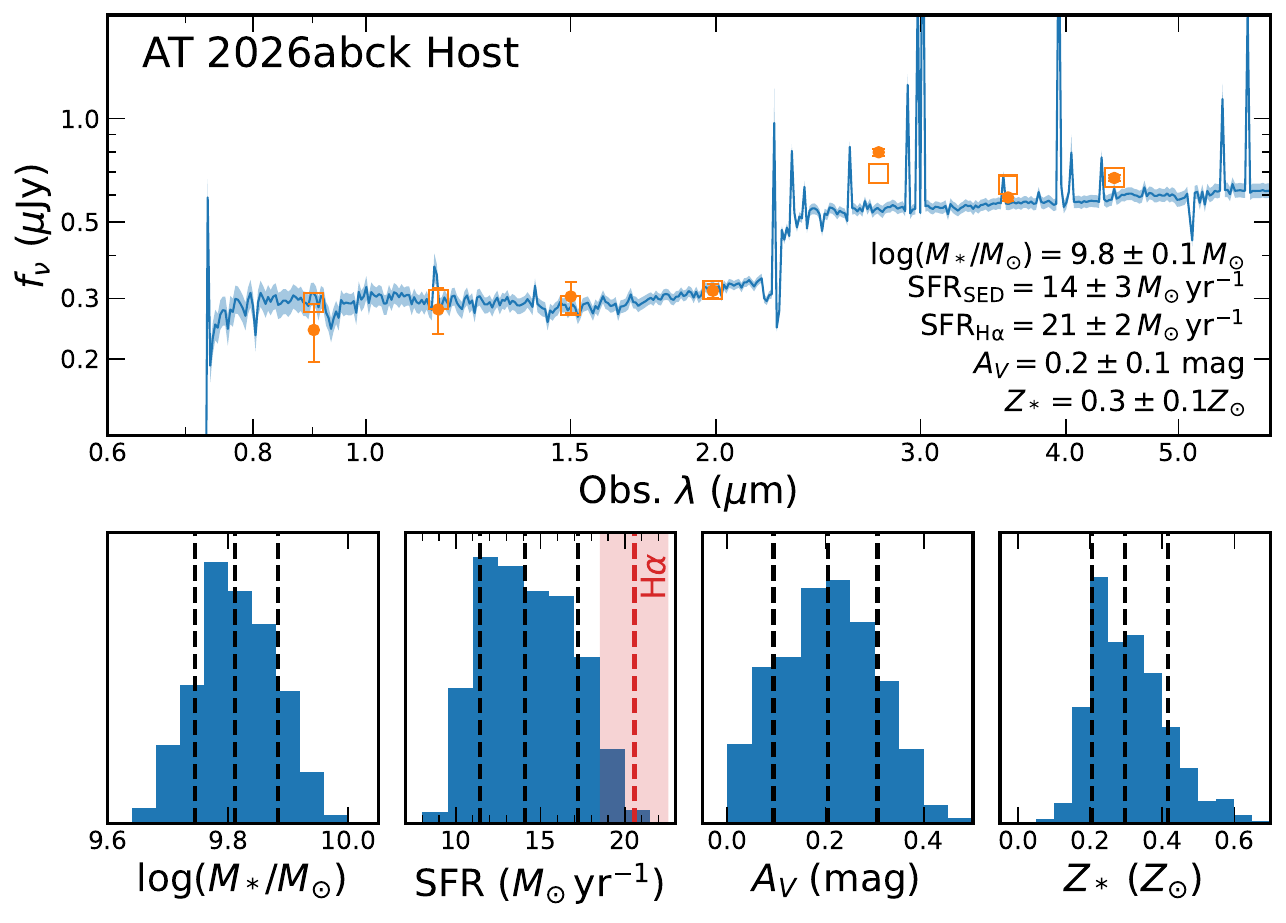}
    \caption{\textbf{Left:} NIRCam F322W2 WFSS of the AT~2026abck host. The top panel shows the full spectrum (blue) and its uncertainty (orange), with \OIII$\lambda5007$ and \Ha\ labeled. The lower panels show the emission-line fits to the \Hb+\OIII\ and \Ha\ windows, with residual shown underneath. Thick curves are the median model and faint curves are drawn from 100 posterior draws. \OIII\ and \Ha\ luminosities are shown in the upper-right corner of the \Ha\ panel. \textbf{Right:} \texttt{Bagpipes} fit to the host spectral energy distribution. The top panel shows the seven-band NIRCam photometry (orange dots) together with the median posterior model spectrum (blue) and its 16th--84th percentile range (shaded). Open orange squares are the median model photometry. Derived properties are quoted at the lower right. The bottom row shows the marginalized posterior distributions of stellar mass ($M_*$), star formation rate (SFR), dust attenuation ($A_V$) and stellar metallicity ($Z_*$), with dashed vertical lines marking the 16th, 50th and 84th percentiles.}
    \label{fig:host_SED}
\end{figure*}

\section{Host Galaxy Properties}\label{sec:host}

Figure~\ref{fig:SN_environment} shows the local environment of AT~2026abck. While two galaxies brighter than $\text{F200W}=28$~mag lie within 3\arcsec\ of the transient, the host association is unambiguous: as shown in Figure~\ref{fig:discovery}, AT~2026abck lies between the two main components of ID101411, with $\sim1.1$~kpc offset from the brighter one.

ID101411 falls within the uniform NEXUS NIRCam WFSS coverage. The F322W2 grism, spanning 2.4--4.0~\micron, covers both \OIII\,$\lambda5007$ and \Ha, whose detection secures a spectroscopic redshift of $z=5.017\pm0.001$ (Figure~\ref{fig:host_SED}). We measure line luminosities of $L_{\rm H\alpha}=3.8 \pm 0.4 \times 10^{42}\, {\rm erg\, s^{-1}}$  and $L_{\rm [O\,III]}=7.7 \pm 0.8 \times 10^{42}\, {\rm erg\, s^{-1}}$. \Hb\ is not significantly detected ($L_{\rm H\beta}=1.5 \pm 1.3 \times10^{42}\, {\rm erg\, s^{-1}}$; $1.2\sigma$), so the Balmer decrement does not constrain the nebular dust attenuation. Adopting the \Ha\ calibration from \citet{Ha_SFR}, we derive an uncorrected star formation rate (SFR) of $21 \pm 2 \,M_{\odot}\,{\rm yr^{-1}}$, which should be treated as a lower limit if dust attenuation is non-negligible.

We measure host fluxes in seven NIRCam filters (F090W, F115W, F150W, F200W, F277W, F356W, and F444W) within an $r=0\farcs625$ aperture, beyond which the curve of growth is flat, using the pre-explosion reference mosaics, except for F277W (Table~\ref{tab:fluxes}). F277W is measured in transient emission-subtracted image using \texttt{galfitm} in Section~\ref{sec:det}. F277W is $\sim1$~mag brighter than F200W owing to strong \OIII\ emission, consistent with the spectroscopic redshift. We model the SED with \texttt{bagpipes} \citep{bagpipes}, adopting a double-power-law star formation history, nebular emission, and the \citet{Calzetti_extinction_curve} attenuation law (Figure~\ref{fig:host_SED}). The fit yields a stellar mass of $\log(M_*/M_\odot)=9.8 \pm 0.1$, sub-solar stellar metallicity ($Z_*=0.3 \pm 0.1 \,Z_\odot$), and low attenuation ($A_V=0.2 \pm 0.1$~mag). The host has a flat UV continuum with $M_{\rm UV}=-21.2 \pm 0.1$~mag. The SED-based SFR, averaged over the past $\sim100$~Myr, is $14 \pm3 \,M_\odot\,{\rm yr^{-1}}$, somewhat lower than the \Ha-based value, which traces the most recent $\lesssim10$~Myr. Both estimates place ID101411 on the $z\approx5$ star-forming main sequence \citep[$\sim14\text{--}57\,M_\odot\,{\rm yr^{-1}}$;][]{SFMS}.

\section{Discussion}\label{sec:discussion}

\subsection{Nature of AT~2026abck}\label{sec:transient_nature}

The properties of AT~2026abck and its host favor an SN origin. The transient lies between the two components of the host, $\sim1.1$ and $\sim1.3$~kpc from the centers of the brighter and fainter components, respectively, both clearly resolved by NIRCam  (Figure~\ref{fig:discovery}). This location disfavors nuclear phenomena such as a tidal disruption event or AGN variability. Such TDE/AGN phenomena would otherwise be difficult to distinguish from a supernova given the hot ($T_{\rm BB}\approx1.7\times10^{4}$~K), compact ($R_{\rm BB}\approx46$~au) photosphere. {Of course, an off-nucleus TDE in an ultra-faint dwarf satellite galaxy cannot be fully ruled out at this point. }

A thermonuclear origin is disfavored by both the SED and the luminosity. The flat UV continuum across rest-frame $\sim2500\text{--}3300$, the F444W non-detection at $\sim7400$~\AA, and the high inferred $T_{\rm BB}$ are inconsistent with SNe~Ia, whose near-UV emission is strongly suppressed by line blanketing from iron-group elements, resulting to much redder SEDs \citep[e.g.,][]{Foley2016}. SNe~Ia rise to peak over $\sim18$ rest-frame days \citep[e.g.,][]{Firth2015}, so at $\lesssim12$~days they are still below their typical peak of $\sim10^{43}\,{\rm erg\,s^{-1}}$. 

The host, ID101411, is an intermediate-mass ($\log(M_*/M_\odot)=9.8$), metal-poor ($Z_*\approx0.3\,Z_\odot$), and mildly dust attenuated ($A_V\approx0.2$~mag) galaxy on the $z\approx5$ star-forming main sequence \citep{SFMS}. Its \Ha-based SFR exceeds the $\sim100$~Myr SED-averaged value, indicating rising or bursty star formation over the past $\lesssim10$~Myr. This environment is dominated by young stellar populations and naturally hosts short-lived massive progenitors. For an SFR of $\sim21\, M_\odot\,{\rm yr^{-1}}$ and $\sim0.01$ CCSNe per $M_{\odot}$ of stars formed \citep[e.g.,][]{MadauDickinson2014, Strolger2015}, the host is expected to produce $\sim0.2$~CCSN {per rest-frame yr.}

Normal stripped-envelope SNe~Ib/c are also disfavored. Because their compact progenitors produce little shock-cooling emission, their early light curves are powered by $^{56}$Ni decay and peak at only $\sim10^{42}$--$10^{43}\,{\rm erg\,s^{-1}}$ with photospheric temperatures of $\lesssim10^{4}$~K \citep[e.g.,][]{Brown2009AJ, Taddia2018}, well below those of AT~2026abck. Hydrogen-poor SNe~Ibn/Icn powered by CSM interaction can be comparably hot and luminous and cannot be excluded photometrically. We focus on SNe~II for the rest of the discussion, noting that the interaction-powered scenarios discussed there have rare hydrogen-poor analogs.

AT~2026abck is hotter and more luminous than normal SNe~II at comparable phases. In the sample of 34 SNe~II with early UV coverage from \citet{Irani2024}, no object exceeds $T_{\rm BB}\approx1.5\times10^{4}$~K beyond $\sim6$~days after explosion, and none reaches $L_{\rm bol}\approx3\times10^{43}\,{\rm erg\,s^{-1}}$ at $\gtrsim4$~days. The closest analog in that sample, SN~2020lfn at 5.4~days, has $T_{\rm BB}\approx1.6\times10^{4}$~K, $R_{\rm BB}\approx7.5\times10^{14}$~cm, and $L_{\rm bol}\approx2.6\times10^{43}\,{\rm erg\,s^{-1}}$, placing AT~2026abck at the extreme upper end of the normal SN~II population. Similarly, the hottest SN~II in the CSP-I sample during the first 10~days reaches only $1.6\times10^{4}$~K \citep{Martinez2022}, and it shows signatures of CSM interaction. Instead, the luminosity of AT~2026abck matches the mean peak luminosity ($\sim3.1\times10^{43}\,{\rm erg\,s^{-1}}$) of SNe~II with the longest-lived signatures of dense confined CSM \citep{JacobsonGalan2024}.

This points to additional early-time power, most plausibly shock-cooling emission from an extended envelope or dense CSM \citep[e.g.,][]{NakarPiro2014, SapirWaxman2017}, or sustained ejecta--CSM interaction as in SNe~IIn. A superluminous SN (SLSN) caught on the rise cannot be excluded given the sub-solar host metallicity \citep[e.g.,][]{Lunnan2014}, although the host is more massive than typical SLSN hosts.

The explosion epoch is constrained by several independent arguments. 
\begin{enumerate}
    \item AT~2026abck is undetected in F200W on 2026 May 28, $\sim71$ observed ($\sim11.8$ rest-frame) days before discovery, to a $3\sigma$ limit of $>28.78$~mag, implying that it brightened by $>1.4$~mag (a factor of $>3.6$ in flux) in the rest-frame near-UV. For an SN~II, which rises to peak within a few days, this non-detection implies that the explosion occurred after the earlier epoch, setting an upper limit of $\sim12$ rest-frame days on the age. If the rise is approximated as exponential, the brightening requires an e-folding time of $\lesssim9$~rest-frame days. This is faster than the rise of the typical SLSNe \citep[e.g., $\sim15\text{--}30$~days;][]{Nicholl2015}, although an SLSN caught early on a steeper rise, $\sim2\text{--}3$~mag below a typical SLSN peak, cannot be ruled out. 

    \item A lower limit on the age follows from the photospheric radius. Assuming free expansion from a negligible initial radius, $R_{\rm BB}$ implies a mean photospheric velocity of $\sim \frac{8100}{(t/10\,{\rm d})} \, {\rm km\, s^{-1}}$, where $t$ is the rest-frame days since explosion. For comparison, $R_{\rm BB}/t$ of normal SNe~II at 4--10~days after explosion has a median value of $\sim10^{4}\, {\rm km\, s^{-1}}$ and does not exceed $\sim2\times10^{4}\, {\rm km\, s^{-1}}$ \citep{Irani2024}, so ages $\lesssim4$~days, which would require $\gtrsim2\times10^{4}\, {\rm km\, s^{-1}}$ , are disfavored.

    \item The temperature provides a complementary constraint: in the \citet{Irani2024} sample, normal SNe~II reach $T_{\rm BB}\gtrsim1.7\times10^{4}$~K only within the first $\sim5$~days, and all have cooled to $\lesssim1.5\times10^{4}$~K by $\sim6$~days. Unless the photosphere is kept hot by CSM interaction, the temperature therefore favors an age of $\lesssim6$~days. 
    
\end{enumerate}

Taken together, these arguments suggest that AT~2026abck was observed $\sim4\text{--}10$ rest-frame days after explosion. The younger end, $\sim5$~days, is preferred if the emission is dominated by shock cooling, whereas an interaction-powered event near peak would be consistent with the 3.5--10~day rise times of SNe~II with dense confined CSM \citep{JacobsonGalan2024}. A large initial radius from an extended envelope or dense CSM would lower the inferred velocities and permit a somewhat younger age. Multi-epoch photometry would both refine the explosion epoch and discriminate among these scenarios. An SN~II would fade rapidly in the rest-frame UV (F150W and F200W) as the photosphere cools, while its rest-frame optical emission (F356W and F444W) settles onto a plateau lasting $\sim100$ rest-frame ($\sim600$ observed) days, producing rapid reddening. In contrast, an SLSN would continue to brighten in all bands over the following tens of rest-frame days. On the other hand, sustained CSM interaction, as in SNe~IIn, would keep the transient hot and UV-bright for longer than in normal SNe~II. {Prompt JWST follow-up observations of AT~2026abck will test these different scenarios. }

\begin{table*}
\centering
\caption{Comparison of SN~Eos and AT~2026abck and their host galaxies.}
\label{tab:SN_comparison}
\begin{tabular}{lcc}
\hline
Property & SN~Eos & AT~2026abck \\
\hline
\multicolumn{3}{c}{Transient} \\
\hline
Redshift & $5.133\pm0.001$ & $5.017\pm0.001$ \\
Classification & Type~IIP (spectroscopic) & CCSN candidate (photometric) \\
Discovery field & Cluster lens, $\mu\approx26\text{--}30$ & Unlensed \\
Rest-frame phase observed & $\lesssim5$~d (HST) and $\sim94\text{--}98$~d (JWST) & $\sim4\text{--}10$~d \\
Early rest-UV absolute magnitude & $\sim-17.3$ to $-18.1$ at $\sim1800$~\AA\tablenotemark{a} & $\sim-19.0$ at $\sim2500$~\AA \\
$T_{\rm BB}$ [K] & $\sim5500$ (late plateau) & $1.7^{+0.3}_{-0.2}\times10^{4}$ (early) \\
Location in host & Coincident with compact UV clump & Between two components; \\
& & $\sim1.1$~kpc from the brighter \\
\hline
\multicolumn{3}{c}{Host galaxy} \\
\hline
$M_{\rm UV}$ [mag] & $-14.4\pm0.3$ & $-21.2\pm0.1$ \\
$\log(M_\ast/M_\odot)$ & $6.5\pm0.4$ & $9.8\pm0.1$ \\
Metallicity [$Z_\odot$] & $\lesssim0.01$ (gas, \OIII/\Hb) & $0.3\pm0.1$ (stellar, SED) \\
$\mathrm{SFR}_{\rm H\alpha}$ [$M_\odot\,{\rm yr}^{-1}$] & $0.11\pm0.02$\tablenotemark{b} & $21\pm2$ \\
$L_{\rm H\alpha}$ [$10^{40}\,{\rm erg\,s^{-1}}$] & $2.1\pm0.4$ & $380\pm40$ \\
$L_{\rm [O\,III] \,5007}$ [$10^{40}\,{\rm erg\,s^{-1}}$] & $<0.49$ ($2\sigma$) & $770\pm80$ \\
\hline
\end{tabular}
\tablecomments{SN~Eos properties are from \citet{SN_Eos} and its host properties from \citet{SN_Eos_host}, corrected for gravitational lensing magnification.}
\tablenotetext{a}{Range over the HST/F110W epochs of both images, demagnified by $\mu=29.9$ (101.2) and 25.5 (101.1).}
\tablenotetext{b}{Derived from $L_{\rm H\alpha}$ with the conversion from \citet{Ha_SFR}.}
\end{table*}

\subsection{Comparison with SN Eos at $z=5.13$}

SN~Eos is the only spectroscopically-confirmed SN at $z\approx 5$ \citep{SN_Eos}. It was discovered by the VENUS program (PID: 6882) as a doubly imaged source behind the galaxy cluster MACS~J1931.8$-$2635, with magnifications of $\mu\approx26\text{--}30$. NIRSpec spectroscopy at $\sim94$ rest-frame days after explosion revealed Balmer P-Cygni profiles, establishing a Type~IIP classification near the end of the plateau, while the weak Fe~II absorption indicates a progenitor metallicity of $Z\lesssim0.1\,Z_\odot$. Archival HST imaging additionally captured SN~Eos in the rest-frame far-UV (1300--1900~\AA) within a few rest-frame days of explosion, which \citet{SN_Eos} modeled as interaction with a dense, confined CSM ($\gtrsim0.2\,M_\odot$ within $10^{15}$~cm, corresponding to a mass-loss rate of $\sim3\times10^{-3}\,M_\odot\,{\rm yr^{-1}}$).

The two transient events are complementary in evolutionary phases. AT~2026abck was observed at $\sim4\text{--}10$ rest-frame days after explosion, a phase that for SN~Eos is sampled only by broadband far-UV photometry and not by JWST. Both objects show early rest-UV emission in excess of that expected from a bare red-supergiant explosion. For SN~Eos, this was modeled with confined CSM, while for AT~2026abck the single-epoch SED favors additional power from either shock cooling of an extended envelope or CSM interaction (Section~\ref{sec:transient_nature}). If CSM interaction is confirmed for AT~2026abck, {e.g., through slowly evolving photospheric temperature and radius with multiband NIRCam imaging or through interaction signatures like high-ionization lines in NIRSpec spectra}, the two events would suggest that the enhanced pre-explosion mass loss common among local SNe~II \citep[e.g.,][]{Forster2018, JacobsonGalan2024} was also present at $z>5$. AT~2026abck is also more luminous at early times: its rest-frame near-UV absolute magnitude ($\sim-19.0$~mag at $\sim2500$~\AA) exceeds that of SN~Eos in its HST epochs ($\sim-17.3$ to $-18.1$~mag at $\sim1800$~\AA). Extrapolating the blackbody fit of AT~2026abck to 1800~\AA\ reduces the difference to $\sim0.5\text{--}1.3$~mag. Nevertheless, the unknown relative phases and the lensing-magnification uncertainty of SN~Eos preclude a precise comparison. Unlike SN~Eos, the discovery of AT~2026abck benefited from the high cadence of NEXUS, demonstrating that even relatively shallow JWST imaging can capture CCSNe at $z\approx5$ during their early, UV-bright phase.

The host galaxies differ far more than the SNe themselves (Table~\ref{tab:SN_comparison}). The host of SN~Eos is an ultra-faint ($M_{\rm UV}=-14.4$~mag), low-mass ($M_\ast\sim10^{6.5}\,M_\odot$) Ly$\alpha$ emitter whose weak \OIII\ emission implies an extremely low gas-phase metallicity ($\lesssim1\%\,Z_\odot$) \citep{SN_Eos_host}. Without lensing it would lie far below the detection limit of any NIRCam survey. In contrast, the host of AT~2026abck is $\sim6.8$~mag brighter in the UV and $\sim3.3$~dex more massive, lies on the $z\approx5$ star-forming main sequence, and has strong \OIII\ emission ($L_{\rm [O\,III]}/L_{\rm H\alpha}\approx2$, compared with $<0.23$ for the SN~Eos host), indicating a substantially more chemically enriched interstellar medium. \citet{SN_Eos_host} showed that, for a CCSN rate per unit SFR independent of metallicity, the host of a CCSN at $z\approx5$ should most likely have $M_{\rm UV}\approx-20$~mag, and that finding one in a host as faint as that of SN~Eos requires either an elevated CCSN rate in metal-poor environments or a rare event. The host of AT~2026abck lies at the bright end of this predicted distribution and is therefore fully consistent with a universal CCSN rate per unit SFR. Together, the two events show that CCSNe at $z>5$ occur across a range of more than three orders of magnitude in host stellar mass and at least an order of magnitude in metallicity. They highlight the potential of high-redshift SNe to probe massive-star formation and death across the full diversity of galaxies within the first billion years.

\section{Conclusions}\label{sec:conclusion}

We reported the discovery of AT~2026abck, a young transient at $z=5.017\pm0.001$ identified in Year-2 observations of the NEXUS JWST Treasury program. Our main conclusions are as follows.

\begin{enumerate}
    \item AT~2026abck was detected on 2026 August 7 at $\sim27.4$~mag in F150W, F200W, and F277W and is undetected in F444W ($>27.87$~mag, $3\sigma$). It is undetected in F200W 71 observed ($\sim11.8$ rest-frame) days earlier, implying a brightening of $>1.4$~mag (a factor of $>3.6$ in flux). 

    \item A blackbody fit to the rest-frame $\sim2500\text{--}7400$~\AA\ photometry yields $T_{\rm BB}\approx1.7\times10^{4}$~K, $R_{\rm BB}\approx7\times10^{14}$~cm, and $L_{\rm bol}\approx3\times10^{43}$~erg~s$^{-1}$. Its off-nucleus location and hot UV-bright SED favor a core-collapse SN origin.

    \item AT~2026abck is hotter and more luminous than typical SNe~II at comparable phases, and its luminosity matches that of SNe~II with dense confined CSM. This points to additional early-time power, most plausibly shock-cooling emission from an extended envelope or CSM interaction. An SLSN on the rise cannot be excluded with current data.
    
    \item Combining the pre-discovery non-detection, the photospheric radius, and the temperature,  our results suggest that AT~2026abck was observed $\sim4\text{--}10$ rest-frame days after explosion, corresponding to a mean photospheric expansion velocity of $\sim 0.8\text{--}2\times10^{4}\, {\rm km\, s^{-1}}$.

    \item Its host is an intermediate-mass ($M_\ast\approx10^{9.8}\,M_\odot$), star-forming galaxy on the $z\approx5$ main sequence, $\sim3.3$~dex more massive than the host of SN~Eos, showing that CCSNe at $z>5$ occur across a wide range of host environments.
    
\end{enumerate}

Detected without lensing magnification, our discovery demonstrates that moderate-depth ($m\approx 28$~mag) but high-cadenced blank-field JWST imaging survey, such as NEXUS, can capture SNe at $z\approx5$ during their early, UV-bright phase. Follow-up observations can distinguish between the proposed scenarios. An SN~II would fade rapidly in the rest-frame UV while settling down onto an optical plateau lasting $\sim600$ observed days, whereas an SLSN would continue to brighten in all bands. Detecting a plateau would require imaging at least $\sim1$~mag deeper than the current F444W limit, and deep NIRSpec spectroscopy would provide a definitive classification. As NEXUS continues its 2-month cadence through 2028, it will build a sample of high-redshift SNe caught at early phases, opening a new window on massive-star death during the first billion years.

\begin{acknowledgments}
Based on observations with the NASA/ESA/CSA James Webb Space Telescope obtained from the Barbara A. Mikulski Archive at the Space Telescope Science Institute, which is operated by the Association of Universities for Research in Astronomy, Incorporated, under NASA contract NAS5-03127. The JWST data presented in this article were obtained
from the Mikulski Archive for Space Telescopes (MAST)
at the Space Telescope Science Institute. Support for Program number JWST-GO-05105 was provided through a grant from the STScI under NASA contract NAS5-03127. L.H. acknowledges support from the STScI grant JWST-AR-05965.
\end{acknowledgments}

\bibliographystyle{aasjournalv7.1}
\bibliography{refs.bib,sn-bibliography.bib}{}

@ARTICLE{NEXUS_transients,
       author = {{Zhuang}, Ming-Yang and {Hu}, Lei and {Pierel}, Justin D.~R. and {Pan}, Zhiwei and {Shen}, Yue and {Burgasser}, Adam J. and {Fan}, Xiaohui and {Ho}, Luis C. and {Li}, Junyao and {Shapley}, Alice E. and {Stone}, Zachary and {Wang}, Feige and {Wang}, Lifan},
        title = "{NEXUS: Transient Searches and First Results from Year One Observations}",
      journal = {arXiv e-prints},
         year = 2026,
        month = sep,
          eid = {arXiv:2609.06985},
        pages = {arXiv:2609.06985},
          doi = {10.48550/arXiv.2609.06985},
archivePrefix = {arXiv},
       eprint = {2609.06985},
 primaryClass = {astro-ph.HE},
       adsurl = {https://ui.adsabs.harvard.edu/abs/2026arXiv260906985Z}
}

@ARTICLE{galfitm,
       author = {{H{\"a}u{\ss}ler}, Boris and {Bamford}, Steven P. and {Vika}, Marina and {Rojas}, Alex L. and {Barden}, Marco and {Kelvin}, Lee S. and {Alpaslan}, Mehmet and {Robotham}, Aaron S.~G. and {Driver}, Simon P. and {Baldry}, I.~K. and {Brough}, Sarah and {Hopkins}, Andrew M. and {Liske}, Jochen and {Nichol}, Robert C. and {Popescu}, Cristina C. and {Tuffs}, Richard J.},
        title = "{MegaMorph - multiwavelength measurement of galaxy structure: complete S{\'e}rsic profile information from modern surveys}",
      journal = {\mnras},
         year = 2013,
        month = mar,
       volume = {430},
       number = {1},
        pages = {330-369},
          doi = {10.1093/mnras/sts633},
archivePrefix = {arXiv},
       eprint = {1212.3332},
 primaryClass = {astro-ph.CO},
       adsurl = {https://ui.adsabs.harvard.edu/abs/2013MNRAS.430..330H}
}

@ARTICLE{SFMS,
       author = {{Popesso}, P. and {Concas}, A. and {Cresci}, G. and {Belli}, S. and {Rodighiero}, G. and {Inami}, H. and {Dickinson}, M. and {Ilbert}, O. and {Pannella}, M. and {Elbaz}, D.},
        title = "{The main sequence of star-forming galaxies across cosmic times}",
      journal = {\mnras},
         year = 2023,
        month = feb,
       volume = {519},
       number = {1},
        pages = {1526-1544},
          doi = {10.1093/mnras/stac3214},
archivePrefix = {arXiv},
       eprint = {2203.10487},
 primaryClass = {astro-ph.GA},
       adsurl = {https://ui.adsabs.harvard.edu/abs/2023MNRAS.519.1526P}
}

@ARTICLE{Ha_SFR,
       author = {{Kennicutt}, Robert C. and {Evans}, Neal J.},
        title = "{Star Formation in the Milky Way and Nearby Galaxies}",
      journal = {\araa},
         year = 2012,
        month = sep,
       volume = {50},
        pages = {531-608},
          doi = {10.1146/annurev-astro-081811-125610},
archivePrefix = {arXiv},
       eprint = {1204.3552},
 primaryClass = {astro-ph.GA},
       adsurl = {https://ui.adsabs.harvard.edu/abs/2012ARA&A..50..531K}
}

@ARTICLE{JADES_transients,
       author = {{DeCoursey}, Christa and {Egami}, Eiichi and {Pierel}, Justin D.~R. and {Sun}, Fengwu and {Rest}, Armin and {Coulter}, David A. and {Engesser}, Michael and {Siebert}, Matthew R. and {Hainline}, Kevin N. and {Johnson}, Benjamin D. and {Bunker}, Andrew J. and {Cargile}, Phillip A. and {Charlot}, Stephane and {Chen}, Wenlei and {Curti}, Mirko and {DeFour-Remy}, Shea and {Eisenstein}, Daniel J. and {Fox}, Ori D. and {Gezari}, Suvi and {Gomez}, Sebastian and {Jencson}, Jacob and {Joshi}, Bhavin A. and {Khairnar}, Sanvi and {Lyu}, Jianwei and {Maiolino}, Roberto and {Moriya}, Takashi J. and {Quimby}, Robert M. and {Rieke}, George H. and {Rieke}, Marcia J. and {Robertson}, Brant and {Shahbandeh}, Melissa and {Strolger}, Louis-Gregory and {Tacchella}, Sandro and {Wang}, Qinan and {Williams}, Christina C. and {Willmer}, Christopher N.~A. and {Willott}, Chris and {Zenati}, Yossef},
        title = "{The JADES Transient Survey: Discovery and Classification of Supernovae in the JADES Deep Field}",
      journal = {\apj},
         year = 2025,
        month = feb,
       volume = {979},
       number = {2},
          eid = {250},
        pages = {250},
          doi = {10.3847/1538-4357/ad8fab},
archivePrefix = {arXiv},
       eprint = {2406.05060},
 primaryClass = {astro-ph.HE},
       adsurl = {https://ui.adsabs.harvard.edu/abs/2025ApJ...979..250D}
}

@ARTICLE{emcee,
       author = {{Foreman-Mackey}, Daniel and {Hogg}, David W. and {Lang}, Dustin and {Goodman}, Jonathan},
        title = "{emcee: The MCMC Hammer}",
      journal = {\pasp},
         year = 2013,
        month = mar,
       volume = {125},
       number = {925},
        pages = {306},
          doi = {10.1086/670067},
archivePrefix = {arXiv},
       eprint = {1202.3665},
 primaryClass = {astro-ph.IM},
       adsurl = {https://ui.adsabs.harvard.edu/abs/2013PASP..125..306F}
}

@ARTICLE{NEXUS_EDR,
       author = {{Zhuang}, Ming-Yang and {Wang}, Feige and {Sun}, Fengwu and {Shen}, Yue and {Li}, Junyao and {Burgasser}, Adam J. and {Fan}, Xiaohui and {Greene}, Jenny E. and {Narayan}, Gautham and {Shapley}, Alice E. and {Yang}, Qian},
        title = "{NEXUS Early Data Release: NIRCam Imaging and WFSS Spectroscopy from the First (Partial) Wide Epoch}",
      journal = {\apjs},
         year = 2026,
        month = feb,
       volume = {282},
       number = {2},
          eid = {54},
        pages = {54},
          doi = {10.3847/1538-4365/ae2d05},
archivePrefix = {arXiv},
       eprint = {2411.06372},
 primaryClass = {astro-ph.GA},
       adsurl = {https://ui.adsabs.harvard.edu/abs/2026ApJS..282...54Z}
}

@ARTICLE{NEXUS_QDR,
       author = {{Zhuang}, Ming-Yang and {Shen}, Yue and {Li}, Junyao and {Pan}, Zhiwei and {Hu}, Lei and {Burgasser}, Adam J. and {Coulter}, David A. and {Greene}, Jenny E. and {Wang}, Feige},
        title = "{NEXUS: Quick Release Notes}",
      journal = {arXiv e-prints},
         year = 2026,
        month = mar,
          eid = {arXiv:2603.04586},
        pages = {arXiv:2603.04586},
          doi = {10.48550/arXiv.2603.04586},
archivePrefix = {arXiv},
       eprint = {2603.04586},
 primaryClass = {astro-ph.IM},
       adsurl = {https://ui.adsabs.harvard.edu/abs/2026arXiv260304586Z}
}

@INPROCEEDINGS{stpsf,
       author = {{Perrin}, Marshall D. and {Sivaramakrishnan}, Anand and {Lajoie}, Charles-Philippe and {Elliott}, Erin and {Pueyo}, Laurent and {Ravindranath}, Swara and {Albert}, Lo{\"\i}c.},
        title = "{Updated point spread function simulations for JWST with WebbPSF}",
    booktitle = {Space Telescopes and Instrumentation 2014: Optical, Infrared, and Millimeter Wave},
         year = 2014,
       editor = {{Oschmann}, Jr., Jacobus M. and {Clampin}, Mark and {Fazio}, Giovanni G. and {MacEwen}, Howard A.},
       series = {Society of Photo-Optical Instrumentation Engineers (SPIE) Conference Series},
       volume = {9143},
        month = aug,
          eid = {91433X},
        pages = {91433X},
          doi = {10.1117/12.2056689},
       adsurl = {https://ui.adsabs.harvard.edu/abs/2014SPIE.9143E..3XP}
}

@ARTICLE{Hu2024,
       author = {{Hu}, Lei and {Wang}, Lifan},
        title = "{Differencing and Coadding JWST Images with Matched Point-spread Function}",
      journal = {\aj},
         year = 2024,
        month = may,
       volume = {167},
       number = {5},
          eid = {231},
        pages = {231},
          doi = {10.3847/1538-3881/ad36cb},
archivePrefix = {arXiv},
       eprint = {2309.09143},
 primaryClass = {astro-ph.IM},
       adsurl = {https://ui.adsabs.harvard.edu/abs/2024AJ....167..231H}
}

@ARTICLE{SExtractor,
       author = {{Bertin}, E. and {Arnouts}, S.},
        title = "{SExtractor: Software for source extraction.}",
      journal = {\aaps},
         year = 1996,
        month = jun,
       volume = {117},
        pages = {393-404},
          doi = {10.1051/aas:1996164},
       adsurl = {https://ui.adsabs.harvard.edu/abs/1996A&AS..117..393B}
}

@ARTICLE{bagpipes,
       author = {{Carnall}, A.~C. and {McLure}, R.~J. and {Dunlop}, J.~S. and {Dav{\'e}}, R.},
        title = "{Inferring the star formation histories of massive quiescent galaxies with BAGPIPES: evidence for multiple quenching mechanisms}",
      journal = {\mnras},
         year = 2018,
        month = nov,
       volume = {480},
       number = {4},
        pages = {4379-4401},
          doi = {10.1093/mnras/sty2169},
archivePrefix = {arXiv},
       eprint = {1712.04452},
 primaryClass = {astro-ph.GA},
       adsurl = {https://ui.adsabs.harvard.edu/abs/2018MNRAS.480.4379C}
}

@ARTICLE{Brown2009AJ,
       author = {{Brown}, Peter J. and {Holland}, Stephen T. and {Immler}, Stefan and {Milne}, Peter and {Roming}, Peter W.~A. and {Gehrels}, Neil and {Nousek}, John and {Panagia}, Nino and {Still}, Martin and {Vanden Berk}, Daniel},
        title = "{Ultraviolet Light Curves of Supernovae with the Swift Ultraviolet/Optical Telescope}",
      journal = {\aj},
         year = 2009,
        month = may,
       volume = {137},
       number = {5},
        pages = {4517-4525},
          doi = {10.1088/0004-6256/137/5/4517},
archivePrefix = {arXiv},
       eprint = {0803.1265},
 primaryClass = {astro-ph},
       adsurl = {https://ui.adsabs.harvard.edu/abs/2009AJ....137.4517B}
}

@ARTICLE{Taddia2018,
       author = {{Taddia}, F. and {Stritzinger}, M.~D. and {Bersten}, M. and {Baron}, E. and {Burns}, C. and {Contreras}, C. and {Holmbo}, S. and {Hsiao}, E.~Y. and {Morrell}, N. and {Phillips}, M.~M. and {Sollerman}, J. and {Suntzeff}, N.~B.},
        title = "{The Carnegie Supernova Project I. Analysis of stripped-envelope supernova light curves}",
      journal = {\aap},
         year = 2018,
        month = feb,
       volume = {609},
          eid = {A136},
        pages = {A136},
          doi = {10.1051/0004-6361/201730844},
archivePrefix = {arXiv},
       eprint = {1707.07614},
 primaryClass = {astro-ph.HE},
       adsurl = {https://ui.adsabs.harvard.edu/abs/2018A&A...609A.136T}
}

@ARTICLE{Firth2015,
       author = {{Firth}, R.~E. and {Sullivan}, M. and {Gal-Yam}, A. and {Howell}, D.~A. and {Maguire}, K. and {Nugent}, P. and {Piro}, A.~L. and {Baltay}, C. and {Feindt}, U. and {Hadjiyksta}, E. and {McKinnon}, R. and {Ofek}, E. and {Rabinowitz}, D. and {Walker}, E.~S.},
        title = "{The rising light curves of Type Ia supernovae}",
      journal = {\mnras},
         year = 2015,
        month = feb,
       volume = {446},
       number = {4},
        pages = {3895-3910},
          doi = {10.1093/mnras/stu2314},
archivePrefix = {arXiv},
       eprint = {1411.1064},
 primaryClass = {astro-ph.HE},
       adsurl = {https://ui.adsabs.harvard.edu/abs/2015MNRAS.446.3895F}
}

@ARTICLE{Foley2016,
       author = {{Foley}, Ryan J. and {Pan}, Yen-Chen and {Brown}, P. and {Filippenko}, A.~V. and {Fox}, O.~D. and {Hillebrandt}, W. and {Kirshner}, R.~P. and {Marion}, G.~H. and {Milne}, P.~A. and {Parrent}, J.~T. and {Pignata}, G. and {Stritzinger}, M.~D.},
        title = "{Ultraviolet diversity of Type Ia Supernovae}",
      journal = {\mnras},
         year = 2016,
        month = sep,
       volume = {461},
       number = {2},
        pages = {1308-1316},
          doi = {10.1093/mnras/stw1440},
archivePrefix = {arXiv},
       eprint = {1604.01021},
 primaryClass = {astro-ph.HE},
       adsurl = {https://ui.adsabs.harvard.edu/abs/2016MNRAS.461.1308F}
}

@ARTICLE{NEXUS,
       author = {{Shen}, Yue and {Zhuang}, Ming-Yang and {Li}, Junyao and {Burgasser}, Adam J. and {Fan}, Xiaohui and {Greene}, Jenny E. and {Narayan}, Gautham and {Shapley}, Alice E. and {Sun}, Fengwu and {Wang}, Feige and {Yang}, Qian},
        title = "{NEXUS: the North ecliptic pole EXtragalactic Unified Survey}",
      journal = {arXiv e-prints},
         year = 2024,
        month = aug,
          eid = {arXiv:2408.12713},
        pages = {arXiv:2408.12713},
          doi = {10.48550/arXiv.2408.12713},
archivePrefix = {arXiv},
       eprint = {2408.12713},
 primaryClass = {astro-ph.GA},
       adsurl = {https://ui.adsabs.harvard.edu/abs/2024arXiv240812713S}
}

@ARTICLE{COSMOS-Web_transients,
       author = {{Fox}, Ori D. and {Rest}, Armin and {Pierel}, Justin D.~R. and {Coulter}, David A. and {Casey}, Caitlin M. and {Kartaltepe}, Jeyhan S. and {Akins}, Hollis B. and {Franco}, Maximilien and {Engesser}, Mike and {Larison}, Conor and {Moriya}, Takashi J. and {Quimby}, Robert M. and {Shuntov}, Marko and {Siebert}, Matthew R. and {DeCoursey}, Christa and {Angulo}, Rodrigo and {DerKacy}, James M. and {Drakos}, Nicole E. and {Egami}, Eiichi and {Finkelstein}, Steven L. and {Flayhart}, Carter and {Fujimoto}, Seiji and {Padilla Gonzalez}, Estefania and {Griggio}, Massimo and {Harish}, Santosh and {Ilbert}, Olivier and {Inayoshi}, Kohei and {Koekemoer}, Anton M. and {Kokorev}, Vasily and {Laigle}, Clotilde and {Lambrides}, Erini and {Larson}, Rebecca L. and {Li}, Xiaolong and {Liu}, Daizhong and {Magdis}, Georgios E. and {McCleary}, Jacqueline E. and {McCracken}, Henry J. and {McMahon}, Nicolas and {McKinney}, Jed and {Moore}, Thomas and {Paquereau}, Louise and {Rhodes}, Jason and {Robertson}, Brant E. and {Sanders}, David B. and {Sanjaripour}, Sogol and {Shukawa}, Koji and {Strolger}, Louis-Gregory and {Toft}, Sune and {Wang}, Qinan and {Williams}, Robert E. and {Zenati}, Yossef},
        title = "{Expanding the High-z Supernova Frontier: ``Wide-area'' JWST Discoveries from the First 2 yr of COSMOS-Web}",
      journal = {\apj},
         year = 2026,
        month = may,
       volume = {1002},
       number = {2},
          eid = {162},
        pages = {162},
          doi = {10.3847/1538-4357/ae5bbf},
archivePrefix = {arXiv},
       eprint = {2601.08931},
 primaryClass = {astro-ph.HE},
       adsurl = {https://ui.adsabs.harvard.edu/abs/2026ApJ..1002..162F}
}

@ARTICLE{SN_Eos,
       author = {{Coulter}, David A. and {Larison}, Conor and {Pierel}, Justin D.~R. and {Fujimoto}, Seiji and {Kokorev}, Vasily and {Allingham}, Joseph F.~V. and {Moriya}, Takashi J. and {Siebert}, Matthew and {Asada}, Yoshihisa and {Bezanson}, Rachel and et al.},
        title = "{A spectroscopically confirmed, strongly lensed, metal-poor Type II supernova at z = 5.13}",
      journal = {arXiv e-prints},
         year = 2026,
        month = jan,
          eid = {arXiv:2601.04156},
        pages = {arXiv:2601.04156},
          doi = {10.48550/arXiv.2601.04156},
archivePrefix = {arXiv},
       eprint = {2601.04156},
 primaryClass = {astro-ph.HE},
       adsurl = {https://ui.adsabs.harvard.edu/abs/2026arXiv260104156C}
}

@ARTICLE{Calzetti_extinction_curve,
       author = {{Calzetti}, Daniela and {Kinney}, Anne L. and {Storchi-Bergmann}, Thaisa},
        title = "{Dust Extinction of the Stellar Continua in Starburst Galaxies: The Ultraviolet and Optical Extinction Law}",
      journal = {\apj},
         year = 1994,
        month = jul,
       volume = {429},
        pages = {582},
          doi = {10.1086/174346},
       adsurl = {https://ui.adsabs.harvard.edu/abs/1994ApJ...429..582C}
}

@ARTICLE{SN_Eos_host,
       author = {{Asada}, Yoshihisa and {Fujimoto}, Seiji and {Allingham}, Joseph F.~V. and {Coulter}, David A. and {Larison}, Conor and {Siebert}, Matthew R. and {Brammer}, Gabriel and {Coe}, Dan and {Dayal}, Pratika and {Fei}, Qinyue and et al.},
        title = "{VENUS: an ultra-faint galaxy hosting the metal-poor type II supernova at $z=5.13$ Witnessing the initial metal enrichment with extremely frequent core-collapse supernovae?}",
      journal = {arXiv e-prints},
         year = 2026,
        month = jul,
          eid = {arXiv:2607.14355},
        pages = {arXiv:2607.14355},
          doi = {10.48550/arXiv.2607.14355},
archivePrefix = {arXiv},
       eprint = {2607.14355},
 primaryClass = {astro-ph.GA},
       adsurl = {https://ui.adsabs.harvard.edu/abs/2026arXiv260714355A}
}

@ARTICLE{MadauDickinson2014,
       author = {{Madau}, Piero and {Dickinson}, Mark},
        title = "{Cosmic Star-Formation History}",
      journal = {\araa},
         year = 2014,
        month = aug,
       volume = {52},
        pages = {415-486},
          doi = {10.1146/annurev-astro-081811-125615},
archivePrefix = {arXiv},
       eprint = {1403.0007},
 primaryClass = {astro-ph.CO},
       adsurl = {https://ui.adsabs.harvard.edu/abs/2014ARA&A..52..415M}
}

@ARTICLE{JacobsonGalan2024,
       author = {{Jacobson-Gal{\'a}n}, W.~V. and {Dessart}, L. and {Davis}, K.~W. and {Kilpatrick}, C.~D. and {Margutti}, R. and {Foley}, R.~J. and {Chornock}, R. and {Terreran}, G. and {Hiramatsu}, D. and {Newsome}, M. and {Padilla Gonzalez}, E. and {Pellegrino}, C. and {Howell}, D.~A. and {Filippenko}, A.~V. and {Anderson}, J.~P. and {Angus}, C.~R. and {Auchettl}, K. and {Bostroem}, K.~A. and {Brink}, T.~G. and {Cartier}, R. and {Coulter}, D.~A. and {de Boer}, T. and {Drout}, M.~R. and {Earl}, N. and {Ertini}, K. and {Farah}, J.~R. and {Farias}, D. and {Gall}, C. and {Gao}, H. and {Gerlach}, M.~A. and {Guo}, F. and {Haynie}, A. and {Hosseinzadeh}, G. and {Ibik}, A.~L. and {Jha}, S.~W. and {Jones}, D.~O. and {Langeroodi}, D. and {LeBaron}, N. and {Magnier}, E.~A. and {Piro}, A.~L. and {Raimundo}, S.~I. and {Rest}, A. and {Rest}, S. and {Rich}, R. Michael and {Rojas-Bravo}, C. and {Sears}, H. and {Taggart}, K. and {Villar}, V.~A. and {Wainscoat}, R.~J. and {Wang}, X.-F. and {Wasserman}, A.~R. and {Yan}, S. and {Yang}, Y. and {Zhang}, J. and {Zheng}, W.},
        title = "{Final Moments. II. Observational Properties and Physical Modeling of Circumstellar-material-interacting Type II Supernovae}",
      journal = {\apj},
         year = 2024,
        month = aug,
       volume = {970},
       number = {2},
          eid = {189},
        pages = {189},
          doi = {10.3847/1538-4357/ad4a2a},
archivePrefix = {arXiv},
       eprint = {2403.02382},
 primaryClass = {astro-ph.HE},
       adsurl = {https://ui.adsabs.harvard.edu/abs/2024ApJ...970..189J}
}

@ARTICLE{Martinez2022,
       author = {{Martinez}, L. and {Bersten}, M.~C. and {Anderson}, J.~P. and {Hamuy}, M. and {Gonz{\'a}lez-Gait{\'a}n}, S. and {Stritzinger}, M. and {Phillips}, M.~M. and {Guti{\'e}rrez}, C.~P. and {Burns}, C. and {Contreras}, C. and {de Jaeger}, T. and {Ertini}, K. and {Folatelli}, G. and {F{\"o}rster}, F. and {Galbany}, L. and {Hoeflich}, P. and {Hsiao}, E.~Y. and {Morrell}, N. and {Orellana}, M. and {Pessi}, P.~J. and {Suntzeff}, N.~B.},
        title = "{Type II supernovae from the Carnegie Supernova Project-I. I. Bolometric light curves of 74 SNe II using uBgVriYJH photometry}",
      journal = {\aap},
         year = 2022,
        month = apr,
       volume = {660},
          eid = {A40},
        pages = {A40},
          doi = {10.1051/0004-6361/202142075},
archivePrefix = {arXiv},
       eprint = {2111.06519},
 primaryClass = {astro-ph.SR},
       adsurl = {https://ui.adsabs.harvard.edu/abs/2022A&A...660A..40M}
}

@ARTICLE{Cooke2012,
       author = {{Cooke}, Jeff and {Sullivan}, Mark and {Gal-Yam}, Avishay and {Barton}, Elizabeth J. and {Carlberg}, Raymond G. and {Ryan-Weber}, Emma V. and {Horst}, Chuck and {Omori}, Yuuki and {D{\'\i}az}, C. Gonzalo},
        title = "{Superluminous supernovae at redshifts of 2.05 and 3.90}",
      journal = {\nat},
         year = 2012,
        month = nov,
       volume = {491},
       number = {7423},
        pages = {228-231},
          doi = {10.1038/nature11521},
archivePrefix = {arXiv},
       eprint = {1211.2003},
 primaryClass = {astro-ph.CO},
       adsurl = {https://ui.adsabs.harvard.edu/abs/2012Natur.491..228C}
}

@ARTICLE{Irani2024,
       author = {{Irani}, Ido and {Morag}, Jonathan and {Gal-Yam}, Avishay and {Waxman}, Eli and {Schulze}, Steve and {Sollerman}, Jesper and {Hinds}, K.-Ryan and {Perley}, Daniel A. and {Chen}, Ping and {Strotjohann}, Nora L. and {Yaron}, Ofer and {Zimmerman}, Erez A. and {Bruch}, Rachel and {Ofek}, Eran O. and {Soumagnac}, Maayane T. and {Yang}, Yi and {Groom}, Steven L. and {Masci}, Frank J. and {Aubert}, Marie and {Riddle}, Reed and {Bellm}, Eric C. and {Hale}, David},
        title = "{The Early Ultraviolet Light Curves of Type II Supernovae and the Radii of Their Progenitor Stars}",
      journal = {\apj},
         year = 2024,
        month = jul,
       volume = {970},
       number = {1},
          eid = {96},
        pages = {96},
          doi = {10.3847/1538-4357/ad3de8},
archivePrefix = {arXiv},
       eprint = {2310.16885},
 primaryClass = {astro-ph.HE},
       adsurl = {https://ui.adsabs.harvard.edu/abs/2024ApJ...970...96I}
}

@ARTICLE{Lunnan2014,
       author = {{Lunnan}, R. and {Chornock}, R. and {Berger}, E. and {Laskar}, T. and {Fong}, W. and {Rest}, A. and {Sanders}, N.~E. and {Challis}, P.~M. and {Drout}, M.~R. and {Foley}, R.~J. and {Huber}, M.~E. and {Kirshner}, R.~P. and {Leibler}, C. and {Marion}, G.~H. and {McCrum}, M. and {Milisavljevic}, D. and {Narayan}, G. and {Scolnic}, D. and {Smartt}, S.~J. and {Smith}, K.~W. and {Soderberg}, A.~M. and {Tonry}, J.~L. and {Burgett}, W.~S. and {Chambers}, K.~C. and {Flewelling}, H. and {Hodapp}, K.~W. and {Kaiser}, N. and {Magnier}, E.~A. and {Price}, P.~A. and {Wainscoat}, R.~J.},
        title = "{Hydrogen-poor Superluminous Supernovae and Long-duration Gamma-Ray Bursts Have Similar Host Galaxies}",
      journal = {\apj},
         year = 2014,
        month = jun,
       volume = {787},
       number = {2},
          eid = {138},
        pages = {138},
          doi = {10.1088/0004-637X/787/2/138},
archivePrefix = {arXiv},
       eprint = {1311.0026},
 primaryClass = {astro-ph.HE},
       adsurl = {https://ui.adsabs.harvard.edu/abs/2014ApJ...787..138L}
}

@ARTICLE{Nicholl2015,
       author = {{Nicholl}, M. and {Smartt}, S.~J. and {Jerkstrand}, A. and {Inserra}, C. and {Sim}, S.~A. and {Chen}, T.-W. and {Benetti}, S. and {Fraser}, M. and {Gal-Yam}, A. and {Kankare}, E. and {Maguire}, K. and {Smith}, K. and {Sullivan}, M. and {Valenti}, S. and {Young}, D.~R. and {Baltay}, C. and {Bauer}, F.~E. and {Baumont}, S. and {Bersier}, D. and {Botticella}, M.-T. and {Childress}, M. and {Dennefeld}, M. and {Della Valle}, M. and {Elias-Rosa}, N. and {Feindt}, U. and {Galbany}, L. and {Hadjiyska}, E. and {Le Guillou}, L. and {Leloudas}, G. and {Mazzali}, P. and {McKinnon}, R. and {Polshaw}, J. and {Rabinowitz}, D. and {Rostami}, S. and {Scalzo}, R. and {Schmidt}, B.~P. and {Schulze}, S. and {Sollerman}, J. and {Taddia}, F. and {Yuan}, F.},
        title = "{On the diversity of superluminous supernovae: ejected mass as the dominant factor}",
      journal = {\mnras},
         year = 2015,
        month = oct,
       volume = {452},
       number = {4},
        pages = {3869-3893},
          doi = {10.1093/mnras/stv1522},
archivePrefix = {arXiv},
       eprint = {1503.03310},
 primaryClass = {astro-ph.SR},
       adsurl = {https://ui.adsabs.harvard.edu/abs/2015MNRAS.452.3869N}
}

@ARTICLE{Gentry2020,
       author = {{Gentry}, Eric S. and {Madau}, Piero and {Krumholz}, Mark R.},
        title = "{Momentum injection by clustered supernovae: testing subgrid feedback prescriptions}",
      journal = {\mnras},
         year = 2020,
        month = feb,
       volume = {492},
       number = {1},
        pages = {1243-1256},
          doi = {10.1093/mnras/stz3440},
archivePrefix = {arXiv},
       eprint = {1912.01141},
 primaryClass = {astro-ph.GA},
       adsurl = {https://ui.adsabs.harvard.edu/abs/2020MNRAS.492.1243G}
}

@ARTICLE{Riess1998,
       author = {{Riess}, Adam G. and {Filippenko}, Alexei V. and {Challis}, Peter and {Clocchiatti}, Alejandro and {Diercks}, Alan and {Garnavich}, Peter M. and {Gilliland}, Ron L. and {Hogan}, Craig J. and {Jha}, Saurabh and {Kirshner}, Robert P. and {Leibundgut}, B. and {Phillips}, M.~M. and {Reiss}, David and {Schmidt}, Brian P. and {Schommer}, Robert A. and {Smith}, R. Chris and {Spyromilio}, J. and {Stubbs}, Christopher and {Suntzeff}, Nicholas B. and {Tonry}, John},
        title = "{Observational Evidence from Supernovae for an Accelerating Universe and a Cosmological Constant}",
      journal = {\aj},
         year = 1998,
        month = sep,
       volume = {116},
       number = {3},
        pages = {1009-1038},
          doi = {10.1086/300499},
archivePrefix = {arXiv},
       eprint = {astro-ph/9805201},
 primaryClass = {astro-ph},
       adsurl = {https://ui.adsabs.harvard.edu/abs/1998AJ....116.1009R}
}

@ARTICLE{Perlmutter1999,
       author = {{Perlmutter}, S. and {Aldering}, G. and {Goldhaber}, G. and {Knop}, R.~A. and {Nugent}, P. and {Castro}, P.~G. and {Deustua}, S. and {Fabbro}, S. and {Goobar}, A. and {Groom}, D.~E. and {Hook}, I.~M. and {Kim}, A.~G. and {Kim}, M.~Y. and {Lee}, J.~C. and {Nunes}, N.~J. and {Pain}, R. and {Pennypacker}, C.~R. and {Quimby}, R. and {Lidman}, C. and {Ellis}, R.~S. and {Irwin}, M. and {McMahon}, R.~G. and {Ruiz-Lapuente}, P. and {Walton}, N. and {Schaefer}, B. and {Boyle}, B.~J. and {Filippenko}, A.~V. and {Matheson}, T. and {Fruchter}, A.~S. and {Panagia}, N. and {Newberg}, H.~J.~M. and {Couch}, W.~J. and {Project}, The Supernova Cosmology},
        title = "{Measurements of {\ensuremath{\Omega}} and {\ensuremath{\Lambda}} from 42 High-Redshift Supernovae}",
      journal = {\apj},
         year = 1999,
        month = jun,
       volume = {517},
       number = {2},
        pages = {565-586},
          doi = {10.1086/307221},
archivePrefix = {arXiv},
       eprint = {astro-ph/9812133},
 primaryClass = {astro-ph},
       adsurl = {https://ui.adsabs.harvard.edu/abs/1999ApJ...517..565P}
}

@ARTICLE{Gelli2024,
       author = {{Gelli}, Viola and {Salvadori}, Stefania and {Ferrara}, Andrea and {Pallottini}, Andrea},
        title = "{Can Supernovae Quench Star Formation in High-z Galaxies?}",
      journal = {\apj},
         year = 2024,
        month = mar,
       volume = {964},
       number = {1},
          eid = {76},
        pages = {76},
          doi = {10.3847/1538-4357/ad23ec},
archivePrefix = {arXiv},
       eprint = {2310.03065},
 primaryClass = {astro-ph.GA},
       adsurl = {https://ui.adsabs.harvard.edu/abs/2024ApJ...964...76G}
}

@ARTICLE{Gal-Yam2019,
       author = {{Gal-Yam}, Avishay},
        title = "{The Most Luminous Supernovae}",
      journal = {\araa},
         year = 2019,
        month = aug,
       volume = {57},
        pages = {305-333},
          doi = {10.1146/annurev-astro-081817-051819},
archivePrefix = {arXiv},
       eprint = {1812.01428},
 primaryClass = {astro-ph.HE},
       adsurl = {https://ui.adsabs.harvard.edu/abs/2019ARA&A..57..305G}
}

@ARTICLE{Rodney2014,
       author = {{Rodney}, Steven A. and {Riess}, Adam G. and {Strolger}, Louis-Gregory and {Dahlen}, Tomas and {Graur}, Or and {Casertano}, Stefano and {Dickinson}, Mark E. and {Ferguson}, Henry C. and {Garnavich}, Peter and {Hayden}, Brian and {Jha}, Saurabh W. and {Jones}, David O. and {Kirshner}, Robert P. and {Koekemoer}, Anton M. and {McCully}, Curtis and {Mobasher}, Bahram and {Patel}, Brandon and {Weiner}, Benjamin J. and {Cenko}, S. Bradley and {Clubb}, Kelsey I. and {Cooper}, Michael and {Filippenko}, Alexei V. and {Frederiksen}, Teddy F. and {Hjorth}, Jens and {Leibundgut}, Bruno and {Matheson}, Thomas and {Nayyeri}, Hooshang and {Penner}, Kyle and {Trump}, Jonathan and {Silverman}, Jeffrey M. and {U}, Vivian and {Azalee Bostroem}, K. and {Challis}, Peter and {Rajan}, Abhijith and {Wolff}, Schuyler and {Faber}, S.~M. and {Grogin}, Norman A. and {Kocevski}, Dale},
        title = "{Type Ia Supernova Rate Measurements to Redshift 2.5 from CANDELS: Searching for Prompt Explosions in the Early Universe}",
      journal = {\aj},
         year = 2014,
        month = jul,
       volume = {148},
       number = {1},
          eid = {13},
        pages = {13},
          doi = {10.1088/0004-6256/148/1/13},
archivePrefix = {arXiv},
       eprint = {1401.7978},
 primaryClass = {astro-ph.CO},
       adsurl = {https://ui.adsabs.harvard.edu/abs/2014AJ....148...13R}
}

@ARTICLE{Cooke2009,
       author = {{Cooke}, Jeff and {Sullivan}, Mark and {Barton}, Elizabeth J. and {Bullock}, James S. and {Carlberg}, Ray G. and {Gal-Yam}, Avishay and {Tollerud}, Erik},
        title = "{Type IIn supernovae at redshift z\raisebox{-0.5ex}\textasciitilde2 from archival data}",
      journal = {\nat},
         year = 2009,
        month = jul,
       volume = {460},
       number = {7252},
        pages = {237-239},
          doi = {10.1038/nature08082},
archivePrefix = {arXiv},
       eprint = {0907.1928},
 primaryClass = {astro-ph.CO},
       adsurl = {https://ui.adsabs.harvard.edu/abs/2009Natur.460..237C}
}

@ARTICLE{Jeon2026,
       author = {{Jeon}, Junehyoung and {Bromm}, Volker and {Venditti}, Alessandra and {Finkelstein}, Steven L. and {Hsiao}, Tiger Yu-Yang},
        title = "{Hunting for the First Explosions at the High-redshift Frontier}",
      journal = {\apj},
         year = 2026,
        month = apr,
       volume = {1001},
       number = {1},
          eid = {3},
        pages = {3},
          doi = {10.3847/1538-4357/ae517d},
archivePrefix = {arXiv},
       eprint = {2601.02469},
 primaryClass = {astro-ph.GA},
       adsurl = {https://ui.adsabs.harvard.edu/abs/2026ApJ..1001....3J}
}

@ARTICLE{Kasen2011,
       author = {{Kasen}, Daniel and {Woosley}, S.~E. and {Heger}, Alexander},
        title = "{Pair Instability Supernovae: Light Curves, Spectra, and Shock Breakout}",
      journal = {\apj},
         year = 2011,
        month = jun,
       volume = {734},
       number = {2},
          eid = {102},
        pages = {102},
          doi = {10.1088/0004-637X/734/2/102},
archivePrefix = {arXiv},
       eprint = {1101.3336},
 primaryClass = {astro-ph.HE},
       adsurl = {https://ui.adsabs.harvard.edu/abs/2011ApJ...734..102K}
}

@ARTICLE{Pierel2025,
       author = {{Pierel}, J.~D.~R. and {Coulter}, D.~A. and {Siebert}, M.~R. and {Akins}, H.~B. and {Engesser}, M. and {Fox}, O.~D. and {Franco}, M. and {Rest}, A. and {Agrawal}, A. and {Ajay}, Y. and {Allen}, N. and {Casey}, C.~M. and {DeCoursey}, C. and {Drakos}, N.~E. and {Egami}, E. and {Faisst}, A.~L. and {Gezari}, S. and {Gozaliasl}, G. and {Ilbert}, O. and {Jones}, D.~O. and {Karmen}, M. and {Kartaltepe}, J.~S. and {Koekemoer}, A.~M. and {Lane}, Z.~G. and {Larson}, R.~L. and {Li}, T. and {Liu}, D. and {Moriya}, T.~J. and {McCracken}, H.~J. and {Paquereau}, L. and {Quimby}, R.~M. and {Rich}, R.~M. and {Rhodes}, J. and {Robertson}, B.~E. and {Sanders}, D.~B. and {Shahbandeh}, M. and {Shuntov}, M. and {Silverman}, J.~D. and {Strolger}, L.~G. and {Toft}, S. and {Zenati}, Y.},
        title = "{Testing for Intrinsic Type Ia Supernova Luminosity Evolution at z > 2 with JWST}",
      journal = {\apjl},
         year = 2025,
        month = mar,
       volume = {981},
       number = {1},
          eid = {L9},
        pages = {L9},
          doi = {10.3847/2041-8213/adb1d9},
archivePrefix = {arXiv},
       eprint = {2411.11953},
 primaryClass = {astro-ph.CO},
       adsurl = {https://ui.adsabs.harvard.edu/abs/2025ApJ...981L...9P}
}

@ARTICLE{RiessLivio2006,
       author = {{Riess}, Adam G. and {Livio}, Mario},
        title = "{The First Type Ia Supernovae: An Empirical Approach to Taming Evolutionary Effects in Dark Energy Surveys from SNe Ia at z>2}",
      journal = {\apj},
         year = 2006,
        month = sep,
       volume = {648},
       number = {2},
        pages = {884-889},
          doi = {10.1086/504791},
archivePrefix = {arXiv},
       eprint = {astro-ph/0601319},
 primaryClass = {astro-ph},
       adsurl = {https://ui.adsabs.harvard.edu/abs/2006ApJ...648..884R}
}

@ARTICLE{Anderson2016,
       author = {{Anderson}, J.~P. and {Guti{\'e}rrez}, C.~P. and {Dessart}, L. and {Hamuy}, M. and {Galbany}, L. and {Morrell}, N.~I. and {Stritzinger}, M.~D. and {Phillips}, M.~M. and {Folatelli}, G. and {Boffin}, H.~M.~J. and {de Jaeger}, T. and {Kuncarayakti}, H. and {Prieto}, J.~L.},
        title = "{Type II supernovae as probes of environment metallicity: observations of host H II regions}",
      journal = {\aap},
         year = 2016,
        month = may,
       volume = {589},
          eid = {A110},
        pages = {A110},
          doi = {10.1051/0004-6361/201527691},
archivePrefix = {arXiv},
       eprint = {1602.00011},
 primaryClass = {astro-ph.GA},
       adsurl = {https://ui.adsabs.harvard.edu/abs/2016A&A...589A.110A}
}

@ARTICLE{Dessart2014,
       author = {{Dessart}, L. and {Gutierrez}, C.~P. and {Hamuy}, M. and {Hillier}, D.~J. and {Lanz}, T. and {Anderson}, J.~P. and {Folatelli}, G. and {Freedman}, W.~L. and {Ley}, F. and {Morrell}, N. and {Persson}, S.~E. and {Phillips}, M.~M. and {Stritzinger}, M. and {Suntzeff}, N.~B.},
        title = "{Type II Plateau supernovae as metallicity probes of the Universe}",
      journal = {\mnras},
         year = 2014,
        month = may,
       volume = {440},
       number = {2},
        pages = {1856-1864},
          doi = {10.1093/mnras/stu417},
archivePrefix = {arXiv},
       eprint = {1403.1167},
 primaryClass = {astro-ph.SR},
       adsurl = {https://ui.adsabs.harvard.edu/abs/2014MNRAS.440.1856D}
}

@ARTICLE{Pessi2023,
       author = {{Pessi}, Thallis and {Anderson}, Joseph P. and {Lyman}, Joseph D. and {Prieto}, Jose L. and {Galbany}, Llu{\'\i}s and {Kochanek}, Christopher S. and {S{\'a}nchez}, Sebastian F. and {Kuncarayakti}, Hanindyo},
        title = "{A Metallicity Dependence on the Occurrence of Core-collapse Supernovae}",
      journal = {\apjl},
         year = 2023,
        month = oct,
       volume = {955},
       number = {2},
          eid = {L29},
        pages = {L29},
          doi = {10.3847/2041-8213/acf7c6},
archivePrefix = {arXiv},
       eprint = {2306.11962},
 primaryClass = {astro-ph.SR},
       adsurl = {https://ui.adsabs.harvard.edu/abs/2023ApJ...955L..29P}
}

@ARTICLE{Ibeling2013,
       author = {{Ibeling}, Duligur and {Heger}, Alexander},
        title = "{The Metallicity Dependence of the Minimum Mass for Core-collapse Supernovae}",
      journal = {\apjl},
         year = 2013,
        month = mar,
       volume = {765},
       number = {2},
          eid = {L43},
        pages = {L43},
          doi = {10.1088/2041-8205/765/2/L43},
archivePrefix = {arXiv},
       eprint = {1301.5783},
 primaryClass = {astro-ph.SR},
       adsurl = {https://ui.adsabs.harvard.edu/abs/2013ApJ...765L..43I}
}

@ARTICLE{Forster2018,
       author = {{F{\"o}rster}, F. and {Moriya}, T.~J. and {Maureira}, J.~C. and {Anderson}, J.~P. and {Blinnikov}, S. and {Bufano}, F. and {Cabrera-Vives}, G. and {Clocchiatti}, A. and {de Jaeger}, T. and {Est{\'e}vez}, P.~A. and {Galbany}, L. and {Gonz{\'a}lez-Gait{\'a}n}, S. and {Gr{\"a}fener}, G. and {Hamuy}, M. and {Hsiao}, E.~Y. and {Huentelemu}, P. and {Huijse}, P. and {Kuncarayakti}, H. and {Mart{\'\i}nez}, J. and {Medina}, G. and {Olivares E.}, F. and {Pignata}, G. and {Razza}, A. and {Reyes}, I. and {San Mart{\'\i}n}, J. and {Smith}, R.~C. and {Vera}, E. and {Vivas}, A.~K. and {de Ugarte Postigo}, A. and {Yoon}, S.-C. and {Ashall}, C. and {Fraser}, M. and {Gal-Yam}, A. and {Kankare}, E. and {Le Guillou}, L. and {Mazzali}, P.~A. and {Walton}, N.~A. and {Young}, D.~R.},
        title = "{The delay of shock breakout due to circumstellar material evident in most type II supernovae}",
      journal = {Nature Astronomy},
         year = 2018,
        month = sep,
       volume = {2},
        pages = {808},
          doi = {10.1038/s41550-018-0563-4},
archivePrefix = {arXiv},
       eprint = {1809.06379},
 primaryClass = {astro-ph.HE},
       adsurl = {https://ui.adsabs.harvard.edu/abs/2018NatAs...2..808F}
}

@ARTICLE{Pierel2024,
       author = {{Pierel}, J.~D.~R. and {Engesser}, M. and {Coulter}, D.~A. and {DeCoursey}, C. and {Siebert}, M.~R. and {Rest}, A. and {Egami}, E. and {Chen}, W. and {Fox}, O.~D. and {Jones}, D.~O. and {Joshi}, B.~A. and {Moriya}, T.~J. and {Zenati}, Y. and {Bunker}, A.~J. and {Cargile}, P.~A. and {Curti}, M. and {Eisenstein}, D.~J. and {Gezari}, S. and {Gomez}, S. and {Guolo}, M. and {Johnson}, B.~D. and {Karmen}, M. and {Maiolino}, R. and {Quimby}, R.~M. and {Robertson}, B. and {Shahbandeh}, M. and {Strolger}, L.~G. and {Sun}, F. and {Wang}, Q. and {Wevers}, T.},
        title = "{Discovery of an Apparent Red, High-velocity Type Ia Supernova at z = 2.9 with JWST}",
      journal = {\apjl},
         year = 2024,
        month = aug,
       volume = {971},
       number = {2},
          eid = {L32},
        pages = {L32},
          doi = {10.3847/2041-8213/ad6908},
archivePrefix = {arXiv},
       eprint = {2406.05089},
 primaryClass = {astro-ph.GA},
       adsurl = {https://ui.adsabs.harvard.edu/abs/2024ApJ...971L..32P}
}

@ARTICLE{Frye2024,
       author = {{Frye}, Brenda L. and {Pascale}, Massimo and {Pierel}, Justin and {Chen}, Wenlei and {Foo}, Nicholas and {Leimbach}, Reagen and {Garuda}, Nikhil and {Cohen}, Seth H. and {Kamieneski}, Patrick S. and {Windhorst}, Rogier A. and {Koekemoer}, Anton M. and {Kelly}, Pat and {Summers}, Jake and {Engesser}, Michael and {Liu}, Daizhong and {Furtak}, Lukas J. and {Polletta}, Maria del Carmen and {Harrington}, Kevin C. and {Willner}, S.~P. and {Diego}, Jose M. and {Jansen}, Rolf A. and {Coe}, Dan and {Conselice}, Christopher J. and {Dai}, Liang and {Dole}, Herv{\'e} and {D'Silva}, Jordan C.~J. and {Driver}, Simon P. and {Grogin}, Norman A. and {Marshall}, Madeline A. and {Meena}, Ashish K. and {Nonino}, Mario and {Ortiz}, Rafael and {Pirzkal}, Nor and {Robotham}, Aaron and {Ryan}, Russell E. and {Strolger}, Lou and {Tompkins}, Scott and {Willmer}, Christopher N.~A. and {Yan}, Haojing and {Yun}, Min S. and {Zitrin}, Adi},
        title = "{The JWST Discovery of the Triply Imaged Type Ia ``Supernova H0pe'' and Observations of the Galaxy Cluster PLCK G165.7+67.0}",
      journal = {\apj},
         year = 2024,
        month = feb,
       volume = {961},
       number = {2},
          eid = {171},
        pages = {171},
          doi = {10.3847/1538-4357/ad1034},
archivePrefix = {arXiv},
       eprint = {2309.07326},
 primaryClass = {astro-ph.GA},
       adsurl = {https://ui.adsabs.harvard.edu/abs/2024ApJ...961..171F}
}

@ARTICLE{Yan2026,
       author = {{Yan}, Haojing and {Sun}, Bangzheng and {Ma}, Zhiyuan and {Wang}, Lifan and {Willmer}, Christopher N.~A. and {Chen}, Wenlei and {Grogin}, Norman A. and {Beacom}, John F. and {Willner}, S.~P. and {Cohen}, Seth H. and {Windhorst}, Rogier A. and {Jansen}, Rolf A. and {Cheng}, Cheng and {Huang}, Jia-Sheng and {Yun}, Min and {Gim}, Hansung B. and {Hammel}, Heidi B. and {Milam}, Stefanie N. and {Koekemoer}, Anton M. and {Hu}, Lei and {Diego}, Jos{\'e} M. and {Summers}, Jake and {D'Silva}, Jordan C.~J. and {Coe}, Dan and {Conselice}, Christopher J. and {Driver}, Simon P. and {Frye}, Brenda and {Marshall}, Madeline A. and {Ortiz}, III, Rafael and {Pirzkal}, Nor and {Robotham}, Aaron and {Ryan}, Jr., Russell E. and {Honor}, Rachel and {O'Brien}, Rosalia and {Fazio}, Giovanni G. and {Adams}, Nathan J. and {Ricotti}, Massimo and {Saikia}, Payaswini and {Hathi}, Nimish P. and {Smith}, Brent and {Holwerda}, Benne W. and {Kelly}, Patrick},
        title = "{PEARLS: 21 Transients Found in the Three-epoch NIRCam Observations in the Continuous Viewing Zone of the James Webb Space Telescope}",
      journal = {\apj},
         year = 2026,
        month = feb,
       volume = {998},
       number = {1},
          eid = {115},
        pages = {115},
          doi = {10.3847/1538-4357/ae2c5d},
archivePrefix = {arXiv},
       eprint = {2506.12175},
 primaryClass = {astro-ph.GA},
       adsurl = {https://ui.adsabs.harvard.edu/abs/2026ApJ...998..115Y}
}

@ARTICLE{Yan2023,
       author = {{Yan}, Haojing and {Ma}, Zhiyuan and {Sun}, Bangzheng and {Wang}, Lifan and {Kelly}, Patrick and {Diego}, Jos{\'e} M. and {Cohen}, Seth H. and {Windhorst}, Rogier A. and {Jansen}, Rolf A. and {Grogin}, Norman A. and {Beacom}, John F. and {Conselice}, Christopher J. and {Driver}, Simon P. and {Frye}, Brenda and {Coe}, Dan and {Marshall}, Madeline A. and {Koekemoer}, Anton and {Willmer}, Christopher N.~A. and {Robotham}, Aaron and {D'Silva}, Jordan C.~J. and {Summers}, Jake and {Nonino}, Mario and {Pirzkal}, Nor and {Ryan}, Russell E. and {Ortiz}, Rafael and {Tompkins}, Scott and {Bhatawdekar}, Rachana A. and {Cheng}, Cheng and {Zitrin}, Adi and {Willner}, S.~P.},
        title = "{JWST's PEARLS: Transients in the MACS J0416.1-2403 Field}",
      journal = {\apjs},
         year = 2023,
        month = dec,
       volume = {269},
       number = {2},
          eid = {43},
        pages = {43},
          doi = {10.3847/1538-4365/ad0298},
archivePrefix = {arXiv},
       eprint = {2307.07579},
 primaryClass = {astro-ph.GA},
       adsurl = {https://ui.adsabs.harvard.edu/abs/2023ApJS..269...43Y}
}

\end{document}